\documentclass[]{elsarticle}
\usepackage{graphicx} 
\usepackage{dcolumn}   
\usepackage{bm}          
\usepackage{amssymb} 
\usepackage{amsmath}
\usepackage{amsthm}
\usepackage{epstopdf}
\usepackage{amsfonts}
\usepackage{verbatim}
\usepackage{feyn}
\usepackage{float}
\usepackage{sidecap}
\usepackage[section]{placeins}
\usepackage{xcolor}
\newcommand{\bbi}{\boldsymbol{i}}
\newcommand{\bj}{\boldsymbol{j}}
\newcommand{\bo}[1]{#1}
\newcommand{\arr}[1]{\overrightarrow{#1}}
\makeatletter
\newcommand{\contraction}[5][1ex]{%
  \mathchoice
    {\contraction@\displaystyle{#2}{#3}{#4}{#5}{#1}}%
    {\contraction@\textstyle{#2}{#3}{#4}{#5}{#1}}%
    {\contraction@\scriptstyle{#2}{#3}{#4}{#5}{#1}}%
    {\contraction@\scriptscriptstyle{#2}{#3}{#4}{#5}{#1}}}%
\newcommand{\contraction@}[6]{%
  \setbox0=\hbox{$#1#2$}%
  \setbox2=\hbox{$#1#3$}%
  \setbox4=\hbox{$#1#4$}%
  \setbox6=\hbox{$#1#5$}%
  \dimen0=\wd2%
  \advance\dimen0 by \wd6%
  \divide\dimen0 by 2%
  \advance\dimen0 by \wd4%
  \vbox{%
    \hbox to 0pt{%
      \kern \wd0%
      \kern 0.5\wd2%
      \contraction@@{\dimen0}{#6}%
      \hss}%
    \vskip 0.2ex%
    \vskip\ht2}}

\newcommand{\contraction@@}[3][0.06em]{%
  \hbox{%
    \vrule width #1 height 0pt depth #3%
    \vrule width #2 height 0pt depth #1%
    \vrule width #1 height 0pt depth #3%
    \relax}}
\newcommand{\Vk}{V}

\begin{document}

\title{Particle Diagrams and Statistics of Many-Body Random Potentials}
\author{Rupert A.~Small}
\ead{Rupert.Small@bristol.ac.uk}
\author{Sebastian M\"uller}
\ead{Sebastian.Muller@bristol.ac.uk}
\address{School of Mathematics, University of Bristol, Bristol BS8 1TW, United Kingdom}
\date{\today}
\begin{abstract}
We present a method using Feynman-like diagrams to calculate the statistical properties of random many-body potentials. This method provides a promising alternative to existing techniques typically applied to this class of problems, such as the method of supersymmetry and the eigenvector expansion technique pioneered in \cite{weid}.  We  use it here to calculate the fourth, sixth and eighth moments of the average level density for systems with  $m$ bosons or fermions that interact through a random $k$-body Hermitian potential ($k \le m$); the ensemble of such potentials with a Gaussian weight is known as the \textit{embedded Gaussian Unitary Ensemble} (eGUE) \cite{mon}. Our results apply in the limit where the number $l$ of available single-particle states is taken to infinity. A key advantage of the method is that it provides an efficient way to identify only those expressions which will stay relevant in this limit. It also provides a general argument for why these terms have to be the same  for bosons and fermions.  The moments are obtained as  sums over ratios of binomial expressions, with a transition from moments associated to a semi-circular level density for $m<2k$ to Gaussian moments in the dilute limit $k \ll m \ll l$. Regarding the form of this transition, we see that as $m$ is increased, more and more diagrams become relevant, with new contributions starting from each of the points $m=2k,3k,\ldots,nk$ for the $2n$-th moment.
\end{abstract}

\maketitle

\section{Introduction}
\noindent Random Matrix Theory (RMT) is the study of random matrices with various symmetry conditions imposed on the matrix entries. With scant warning this young theory has permeated nearly every area of modern physics and even number theory \cite{mehta, guhr, akemann}.
In quantum physics, random matrices can be used to model the behaviour of Hamiltonians or scattering matrices, and many statistical properties of chaotic quantum systems have been found to agree with the appropriate predictions from RMT \cite{akemann, bgs}.
In recent decades attempts have been made to further refine what has become the canonical theory, by considering symmetries which allow the matrix representations of quantum potentials to impose $k$-body interactions between the particles in a system containing $m$ particles ($k \le m$). Here the main new feature is that such an interaction, when applied to one of the states with $m$ particles, will annihilate $k$ particles and create $k$ particles in (possibly) different single-particle states. This means that matrix elements between many-particle states that differ by more than $k$ occupied single-particle states will necessarily be zero. Canonical RMT, containing no such restrictions, can be associated with the
case $k=m$. The case $k=1$ describes random single-particle potentials and hopping terms. Although the most common interactions have $k=2$ it remains of great interest to determine the statistics of such interactions for the whole physically relevant domain $k \le m$.

The appropriate generalization of canonical RMT  involves \textit{embedding} the \textit{k}-body potential into the \textit{m}-particle state space creating what has become known as the embedded ensembles. The embedded ensembles, first introduced by Mon and French \cite{mon} in 1975, gave physicists a powerful framework for studying many-body interactions using random matrix theory. (See \cite{weidreview,kota} for reviews, and \cite{tbre1,tbre2} for the related two-body random ensemble.) In particular, the embedded Gaussian Unitary Ensemble of random matrices (eGUE) represents the Hamiltonian of non time-reversal invariant quantum systems of $m$ particles interacting under the force of a $k$-body potential, so called because the potential is a sum of interaction terms between $k$-tuples of particles. In addition, many-body Hamiltonians of a similar form are used independently to study the statistics of quantum spin chains, spin glasses and (hyper)graphs \cite{hartmann,huw,schroder} and recent developments point to a convergence of some statistical properties  between these models \cite{schroder,small, nakata}.

In one of the main contributions to this area Benet, Rupp and Weidenm\"uller \cite{weid}  showed how a process of eigenvector expansions could be used to calculate certain statistical properties of $k$-body potentials, in particular the fourth moment of the average level density. Though a great advance, the eigenvector expansion method is complex to implement, and it remains unclear if it can practically be used to calculate moments higher than the fourth. The method of supersymmetry, also used in \cite{weid} to investigate the fourth moment, is accompanied by technical difficulties in the loop expansion, and does not allow one to access the regime $m \geq 2k$. A further technique used to treat embedded ensembles is the trace propagation method \cite{kota}. Using a new method however, which utilises Feynman-like diagrams to simplify calculations, we will show that it becomes possible to calculate the fourth, sixth and eighth moments for embedded ensembles in a straightforward way. The method, which we will call the method of \textit{particle diagrams}, is designed to probe the order of magnitude of combinatorial expressions prior to calculating them explicitly. We will specifically be interested in the case where, in correspondence with many physical systems, the number of available single-particle states $l$ is taken to infinity. In this limit estimating the order of magnitude provides a sufficient excuse not to calculate certain terms at all, since we can foretell using particle diagrams that they will not survive in this asymptotic regime. Hence by applying the method of particle diagrams one is in effect washing out much of the complexity of the problem, with enough details remaining to yield limiting statistics.

We will present this technique in detail here, significantly extending our previous rapid communication \cite{small}. First, we will introduce the method using the fourth moment as an example. Afterwards we will proceed to the sixth and eighth moments, using a further methodological development that involves studying closed loops on particle diagrams.

A general feature of our results will be a gradual transition from the moments of a semi-circular average level density for $m<2k$ to Gaussian moments for large $m$, or more precisely in the dilute limit $k\ll m\ll l$. Reassuringly, the semi-circular regime contains the case $m=k$ corresponding to canonical RMT. The Gaussian behaviour is in line with early observations in \cite{mon} as well as results for quantum spin chains and graphs \cite{hartmann,huw,schroder}. Our general result, interpolating between these two regimes, will be a sum over binomial expressions depending on $m$ and $k$. We will see that as $m$
is increased the terms responsible for the transition in the $2n$-th moment (at least for $n=2,3,4$) start to contribute immediately from $m=2k, 3k, \ldots, nk$. Correspondingly, if we increase $k$ for fixed $m$ we have
fewer and fewer terms contributing, with contributions vanishing after $k=\frac{m}{n}, \frac{m}{n-1}, \ldots \frac{m}{2}$; after reaching  $k=\frac{m}{2}$ all subsequent values of the moments are semi-circular. This form of the transition is consistent with \cite{weid} for the fourth moment but differs from the case of spin glasses where, translated into our notation, the transition occurs when $m$ is of the order of $k^2$ \cite{schroder}.

\section{Particle diagrams and the embedded GUE}\label{sec:egue}
\subsection{The embedded GUE}
\label{embedded_subsection}
 In this section we reproduce a known result for the normalised fourth moment, or kurtosis, of the level density for the eGUE using a simpler alternate calculation to that found in \cite{weid}. This will serve as a platform for introducing both the basic definitions of the field and the new methodology involving particle diagrams which this paper sets out to explain.  We will consider states of $m$ fermions in a system with $m \ll l$ single-particle states, all interacting under the action of a \textit{k}-body potential ($k \le m$) with an identical gaussian \textit{p.d.f} determining its independent entries. We will neglect spin. Our results will thus apply if the available states are either not distinguished by a spin degree of freedom; or alternatively if the modelled Hamiltonians have no specific features related to spin (such as coupling terms involving a spin operator) so that the interactions can be randomised in the same way regardless of whether the interacting states are distinguished by spin or other degrees of freedom. A generalization to bosons will be given in subsection \ref{bosons}.

\begin{figure}[t]
\centering
\includegraphics[scale=.5]{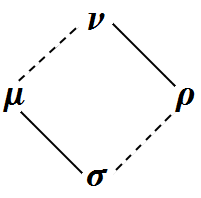}
\caption{\label{fig:square} Particle diagram of  $A_{\mu\nu\rho\sigma}=\langle \mu|a_{\bj}^{\dag} a_{\bbi}  |\sigma \rangle \langle \rho|a_{\bbi}^{\dag}  a_{\bj} |\nu \rangle$
as in  (\ref{eq:ee46}).  Each bond between compound states represents a set of single-particle states shared by both of the compound states.}
\end{figure}

The single-particle creation and annihilation operators are $a_j^{\dag}$ and $a_j$ respectively with $j = 1, \ldots, l$ and we define a shorthand notation for products of these with the abbreviation $\bj=(j_1,\ldots,j_k)$, $a_{\bj}=a_{j_k}\ldots a_{j_1}$ (similarly  for $\bbi)$. A useful corollary of this is
\begin{equation}\label{eq:ee30}a_{\bj}^{\dag} = a_{j_1}^{\dag} \ldots a_{j_k}^{\dag}\end{equation}
We will write the orthonormal $m$-particle states as $|\mu\rangle, |\nu\rangle, |\rho\rangle$, etc, where each state takes the form $a_{{j_m}}^{\dag}\ldots a_{{j_1}}^{\dag}|0\rangle$ with $|0\rangle$ denoting the vacuum state and the restriction $1\leq j_1 < j_2 < \ldots < j_m\leq l$. The $k$-body potential \cite{orland} is given by
\begin{equation}\label{eq:ee27}{V}_k = \sum_{{{1 \le j_1 < \ldots < j_k \le l} \atop {1 \le i_1 < \ldots < i_k \le l}}} {v}_{j_1 \ldots j_k;i_1 \ldots i_k} a_{j_1}^{\dag} \ldots a_{j_k}^{\dag} a_{i_k} \ldots a_{i_1}\end{equation}
which can now be abbreviated to
\begin{equation}\label{eq:egue01}{V}_k = \sum_{\bj, \bbi}v_{\bj\bbi} a_{\bj}^{\dag}  a_{\bbi}\;.\end{equation}
As we are considering fermions all  $m$-particle states are assumed to contain only non-repeating single-particle states and since the number of allowed single-particle states is $l$, the dimension of the $m$-body state space is $N={l\choose m}$. For the embedded GUE ensemble presently under consideration the only symmetry condition on the potential is that it be Hermitian
\begin{align}\label{eq:ee33}   \langle \mu|V_k|\nu \rangle = \langle \nu|V_k|\mu \rangle ^{*} \end{align}
for all $\mu, \nu$ which implies that
\begin{align}\label{eq:ee34}\sum_{\bj, \bbi} {v}_{\bj\bbi} \langle\mu|a_{\bj}^{\dag} a_{\bbi}|\nu\rangle &= \sum_{\bj, \bbi} {v}_{\bbi\bj}^{*} \langle\mu|a_{\bj}^{\dag} a_{\bbi}|\nu\rangle\;.\nonumber\\
\end{align}
Matching coefficients gives
\begin{align}\label{eq:ee35} {v}_{\bj\bbi}^{*} = {v}_{\bbi\bj}\;.\end{align}
As in canonical RMT (where $k=m$) we suppose that matrix elements not related by Hermitian symmetry are uncorrelated \textit{i.i.d} complex Gaussian random variables with mean zero and variance $v_o^2$ and without loss of generality we take $ v_o^2 = 1$. Hence for uncorrelated $v_{ \bj \bbi}$ and $v_{\bj' \bbi'}$ one has $\overline {v_{\bj\bbi} v_{\bj'\bbi'} } = 0$ whereas for $v_{\bj \bbi} = v_{\bj{'}\bbi{'}}^*$ the average instead becomes unity so that
\begin{equation}\label{average}
\overline {v_{\bj \bbi} v_{\bj{'} \bbi{'}}} = \delta_{\bj \bbi{'}}\delta_{\bbi\bj^{'}}\;.\end{equation}
The analogous relation for matrix elements of $V_k$ is
\begin{align}\label{eq:ee46}A_{\mu\nu\rho\sigma} :=& \overline{\langle \mu|V_k|\sigma \rangle\langle \rho|V_k|\nu \rangle}\nonumber\\=&\overline{v_{\bj\bbi}v_{\bj'\bbi'}}
\langle\mu| a_{\bj}^\dagger a_{\bbi}|\sigma\rangle\langle\rho|a_{\bj'}^\dagger a_{\bbi'}|\nu\rangle =
\langle \mu|a_{\bj}^{\dag}  a_{\bbi} |\sigma \rangle \langle \rho|a_{\bbi}^{\dag}a_{\bj} |\nu \rangle.\end{align}
Here summation over repeated indices $\bbi,\bj,\bbi',\bj'$ is implied. The quantity $A_{\mu\nu\rho\sigma}$ will be a crucial ingredient for the following calculations. (Note the unusual ordering of indices in line with the notation of \cite{weid}.)
For $A_{\mu\nu\rho\sigma}$ to be non-vanishing $|\sigma\rangle$ and $|\rho\rangle$ must both contain the $k$ single-particle states included in $\bbi$, and  $|\mu\rangle$ and $|\nu\rangle$ must both contain the $k$ single-particle states included in $\bj$. In addition $a_{\bbi}|\mu\rangle$ and $a_{\bj}|\sigma\rangle$ have to contain the same single-particle states implying that $|\mu\rangle$ and $|\sigma\rangle$ coincide in the $m-k$ single-particle states not included in $\bbi$ or $\bj$, and the same applies to $|\rho\rangle$ and $|\nu\rangle$. These relations are illustrated by  {\it particle diagrams} as in Fig. \ref{fig:square} where solid bonds $~\feyn{f}~$ connect states sharing $m-k$ single-particle states and dashed bonds $~\feyn{h}~$ connect many-particle states sharing $k$ single-particle states. Note that in the figure the overlaps indicated by neighbouring bonds are disjoint, e.g., the single-particle states of $\bbi$ form the overlap $\sigma\feyn{h}\rho $ but are excluded from the overlap $\mu\feyn{f}\sigma$ because the state $|\sigma\rangle$ contains $m$ non-repeated single-particle states.
Furthermore note that every single-particle state included in $|\sigma\rangle$ is also included in either $|\rho\rangle$ (due to the overlap $\sigma\feyn{h}\rho$) or in $|\mu\rangle$ (due to the overlap $\mu\feyn{f}\sigma$). A similar argument applies to $|\nu\rangle$. As the sizes of all states are equal this implies that the union of the many-particle states $|\sigma\rangle$ and $|\nu\rangle$ coincides with the union of $|\mu\rangle$ and $|\rho\rangle$.
Fixing either $|\sigma\rangle$ and $|\nu\rangle$, or $|\mu\rangle$ and $|\rho\rangle$
thus determines all participating single-particle states.

\subsection{Moments}
 We are now interested in moments of the level density $\rho(E)$ of $V_k$. These moments can be expressed through traces of the powers of $V_k$ using the fact that
$\int \rho(E)E^p dE={\rm tr} \overline{V_k^p}$. After ensemble averaging the odd moments trivially vanish, because for odd $p$ positive and negative contributions mutually cancel. Using the second moment for normalization we thus have to evaluate the $2n$-th moments given by\begin{equation}\label{eq:moments}
\beta_{2n} = \frac{\frac{1}{N}\mathrm{tr}({\overline{V^{2n}_k}})}{\left(\frac{1}{N}\mathrm{tr}({\overline{V^2_k}})\right)^n}
.\end{equation}
We start with the fourth moment, or kurtosis
\begin{equation}\label{eq:ee48}\kappa = \frac{\frac{1}{N}\mathrm{tr}(\overline{V_{k}^4})}{\left(\frac{1}{N}\mathrm{tr}(\overline{V_{k}^2})\right)^2}\;.\end{equation}
The terms in the numerator and in the denominator of $\kappa$ can be evaluated using Wick's law which expresses the Gaussian average as a sum over all ways  to draw contraction lines between the factors $V_k$,
\begin{align}
\mathrm{tr}(\overline{V_k^2})&=\langle\mathrm{tr} \contraction[2ex]{}{V_k}{}{V_k}{}
V_k V_k\rangle\\
\label{trV4}
\mathrm{tr}(\overline{V_k^4})&=
\langle\mathrm{tr}
\contraction[2ex]{}{V_k}{}{V_k}
\contraction[2ex]{V_kV_k}{V_k}{}{V_k}
V_k V_k V_k V_k\rangle
+ \langle
\mathrm{tr}
\contraction[4ex]{}{V_k}{V_kV_k}{V_k}
\contraction[2ex]{V_k}{V_k}{}{V_k}
V_k V_k V_k V_k\rangle
+\langle
\mathrm{tr}
\contraction[2ex]{}{V_k}{V_k}{V_k}
\contraction[4ex]{V_k}{V_k}{V_k}{V_k}
V_k V_k V_k V_k\rangle.
\end{align}
Here the brackets $\langle.\rangle$ are used as an alternative notation for the ensemble average.
We note that the first two contributions to (\ref{trV4}) coincide due to cyclic invariance of the trace. We can now write out the traces of powers in terms of matrix elements. Wick's theorem then allows us to compute the averages of contracted matrix elements as if the remaining elements were absent. The result can be expressed in terms of (\ref{eq:ee46}), leading to
\begin{align}\label{eq:ee49}
\mathrm{tr}(\overline{V_{k}^2}) =&  A_{\mu\mu\rho\rho}\\
\label{eq:ee52}
\mathrm{tr}(\overline{V_{k}^4}) =&  2 A_{\mu\nu\rho\rho}A_{\nu\mu\sigma\sigma} + A_{\mu\nu\rho\sigma}A_{\sigma\mu\nu\rho}
\end{align}
with the summations over repeated indices $\mu, \nu, \rho, \sigma$ implicit.
The required expressions for higher moments of the level density will look similar but with more factors $A$, and (unless further simplification is possible) each many-particle subscript has to appear twice.

In (\ref{eq:ee49}) given the restrictions from Fig. \ref{fig:square} we have to sum over all $|\mu\rangle$ and $|\rho\rangle$ sharing $m-k$ single-particle states. There are $N = {l\choose m}$ states in the sum over all possible $|\mu\rangle$, ${m\choose m-k}$ ways to choose the overlap with $|\rho\rangle$, and ${l-(m-k)\choose k}$ ways to choose the rest of $|\rho\rangle$. Hence the result is
\begin{equation}\label{eq:arg3} \mathrm{tr}(\overline{V_{k}^2}) = {l \choose m}{m \choose k}{{l-m+k}\choose k}.\end{equation}

\subsection{Large-$l$ asymptotics and arguments}\label{arguments}
In this paper we are interested only in the limit $l\to\infty$.
We can thus use Stirling's formula $l!\sim \sqrt{2\pi l}\left(\frac{l}{e}\right)^l$ (with $\sim$ denoting asymptotic equality in the limit $l\to\infty$)
to approximate
\begin{equation}\label{factorial}
{{l-a} \choose b}\sim\frac{1}{b!}\left(\frac{l}{e}\right)^b
\end{equation}
for fixed $a,b\ll l$ and thus
\begin{align}\label{divisor}
\frac{1}{N}\mathrm{tr}(\overline{V_k^2})\sim\frac{1}{k!}{m\choose k}\left(\frac{l}{e}\right)^k.
\end{align}
For traces of higher powers of $V_k$ we will similarly encounter binomial or multinomial factors of the type $\prod_n{{l - a_n} \choose b_n}^{i_n}$ or ${l-a\choose b_1 \; b_2 \; b_3\; \ldots}$.
These factors count ways to select from all $l$ single-particle state (or a number already reduced by $a_n$ or $a$) sets of $b_1,b_2,b_3,\ldots$ states to be included in the relevant many-particle states appearing as subscripts of the factors $A$.
To be able to handle the limit $l\to\infty$ efficiently, we  define the {\it} argument $\arg$ of an $l$-dependent binomial expression as its dominating power in $l$, leading to
\begin{equation}
\arg\left [\prod_n{{l - a_n} \choose b_n}^{i_n}\right
] = \sum_n i_n b_n
\end{equation}
and
\begin{equation}
\label{eq:tr4A}
\arg{l-a\choose b_1 \; b_2 \; b_3\; \ldots}=\sum_nb_n.
\end{equation}
Physically the argument thus gives the overall number of single-particle states included in the many-particle states. The argument of the numerator in the definition of the moments (\ref{eq:moments})
will always turn out to be smaller or equal to the argument of the denominator.
This implies a huge simplification: For every contribution to the moments we only have to consider those choices for $b_1,b_2,b_3,\ldots$ that {\it maximise the argument} of the numerator, and thus the number of participating single-particle states. In case even these choices lead to an argument smaller than the one in the denominator, the corresponding contribution can be ignored completely. If the argument is equal the case of maximal argument in the numerator is sufficient to fully determine the contribution to the moments in the limit $l\to\infty$ up to corrections of order $1/l$ or smaller. This approach washes out much of the complexity attendant to the study of embedded ensembles.

For later reference it is useful  to note that  due to (\ref{divisor}) the denominator of the $2n$-th moment (\ref{eq:moments}) has the argument $nk$. And since $\arg N=\arg{l\choose m}=m$ the contributions to $\mathrm{tr}\overline{V_k^{2n}}$
thus have to involve $m+nk$ single-particle states in order to be relevant in the limit $l\to\infty$. Also for later reference, we want to state the asymptotic form of the most common type of multinomial to be found below.
Using (\ref{factorial}) one can show that
\begin{equation}
\label{multi1}
{l\choose \underbrace{k\;\ldots\;k}_{\mbox{$n+i$ terms}}m-ik}\sim\frac{1}{m!k!^n}\prod_{j=0}^{i-1}{m-jk\choose
k}\left(\frac{l}{e}\right)^{m+nk}
\end{equation}
which combines with (\ref{divisor}) and the asymptotics of $N$ to give
\begin{equation}
\label{multi2}
\frac{\frac{1}{N}{l\choose{k\;\ldots\;k\;m-ik}}}{\left(\frac{1}{N}\mathrm{tr}(\overline{V_k^2})\right)^n}\sim\frac{\prod_{j=0}^{i-1}{m-jk\choose
k}}{{m\choose k}^n}\;
\end{equation}
where $k$ is again repeated $n+i$ times.

\subsection{Identical subscripts}\label{identical}
For the term $A_{\mu\nu\rho\rho}A_{\nu\mu\sigma\sigma}$ in (\ref{eq:ee52}) and many terms in higher-order moments there is a further simplification, due to the fact that at least one factor $A$ has coinciding first and second, or third and fourth subscripts representing neighbouring many-particle states in Fig. \ref{fig:square}.
This situation always arises when Wick contractions connect two neighbouring factors $V_k$, so that either the $\sigma$ and $\rho$ or $\mu$ and $\nu$ in the arising factor $A_{\mu\nu\rho\sigma} = \overline{\langle \mu|V_k|\sigma \rangle\langle \rho|V_k|\nu \rangle}$ coincide.
In this case one can show that a contribution to the moments is obtained only if the other two subscripts of this $A$ coincide as well. Taking the factor $A_{\mu\nu\rho\rho}$ as an example we have to consider the diagram of Fig. \ref{fig:square} with $|\sigma\rangle=|\rho\rangle$. The many-particle state $|\sigma\rangle=|\rho\rangle$
then contains $k$ states forming the bond $\sigma\feyn{h}\rho$ and $m-k$ further states. Given that each state is the union of the two attached bonds this means that the remaining $m-k$ states must form the bonds $\mu\feyn{f}\sigma$ and $\rho\feyn{f}\nu$, which have to be identical. Then the states $|\mu\rangle$ and $|\nu\rangle$ share both the latter bond and the bond $\mu\feyn{h}\nu$. As a consequence they have to coincide as well. Whenever we encounter a contribution like this we thus have to (i) evaluate the remaining product of $A$'s for the case $|\mu\rangle=|\nu\rangle$ and then (ii) take into account the additional freedom of choice due to the factor $A_{\mu\mu\rho\rho}$. (i) leads to a term already contributing to a lower-order moment, with the index $|\mu\rangle=|\nu\rangle$ appearing twice and no occurrences of $|\rho\rangle$.
For (ii) one uses that $|\rho\rangle=|\sigma\rangle$ and $|\mu\rangle=|\nu\rangle$ have to share $m-k$ states, and there are ${m \choose m-k}={m\choose k}$ ways to select these out of the $m$ states in $|\mu\rangle$. The remaining
$k$ states in $|\rho\rangle$ can be chosen among all $l-(m-k)$ states not already included. Altogether step (ii) thus leads to a factor ${m \choose k}{l-m+k \choose k}$. Due to (\ref{eq:arg3}) this  compensates one of the factors $\frac{1}{N}\mathrm{tr} \overline{V_k^2}$ in the denominator (even for finite $l$!). We thus get back a diagram contributing to a lower moment, and if the normalization is taken into account we see that this diagram gives the same contribution as the original one. If the result of (i) again involves identical subscripts in a factor $A$ this procedure can be applied recursively leading to a diagram of an even lower order. Applying this idea to the first term in the fourth moment and afterwards invoking (\ref{eq:arg3}) we obtain
\begin{equation}
\label{reduced_example}
2\frac{\frac{1}{N}A_{\mu\nu\rho\rho}A_{\nu\mu\sigma\sigma}}{\left(\frac{1}{N}\mathrm{tr}(\overline{V_k^2})\right)^2}=2\frac{\frac{1}{N}A_{\mu\mu\sigma\sigma}}{\frac{1}{N}\mathrm{tr}(\overline{V_k^2})}=2.
\end{equation}
At the level of contraction lines, the present procedure amounts to removing two neighbouring factors $V_k$ connected by contraction lines.
Recursive application, if possible, allows to remove more of these factors and lines. If none of the contraction lines intersect, all $V_k$ can be removed in this way and as in (\ref{reduced_example}) the final result is just the multiplicity factor of the diagram.

For an interpretation in terms of particle diagrams one can draw the states identifying $|\mu\rangle$ and $|\nu\rangle$  and then represent the overlap of $m-k$ states between  by a ``tail'' $\mu\feyn{f}\rho$.  Our result  implies that such tails can be removed from the diagram without changing its normalised contribution.
\begin{figure}[t]
\centering
\includegraphics[scale=.5]{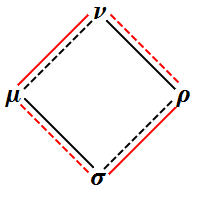}
\caption{\label{fig:q_term}\label{fig:standard} Particle diagram  $A_{\mu\nu\rho\sigma}A_{\sigma\mu\nu\rho}$.
The inner (black) bonds show sets of single-particle states that have to be shared between the compound states in order to have non-zero $A_{\mu\nu\rho\sigma}$ (as in Fig. \ref{fig:square}), the outer
(red) bonds arise from the factor   $A_{\sigma\mu\nu\rho}$.}
\end{figure}

\subsection{Diagram  $A_{\mu\nu\rho\sigma}A_{\sigma\mu\nu\rho}$}
\label{standard}
The particle diagram for $A_{\mu\nu\rho\sigma}A_{\sigma\mu\nu\rho}$ is depicted in Fig. \ref{fig:standard}. For this diagram non-zero contributions arise if the indices of $A_{\mu\nu\rho\sigma}$ obey the same restrictions as introduced earlier with reference to (\ref{eq:ee46}) and depicted again by the inner (black) bonds in Fig.  \ref{fig:standard}. The analogous restrictions to have non-vanishing $A_{\sigma\mu\nu\rho}$ are depicted by the outer (red) bonds.  We will have occasion to use the combined diagram throughout this paper in calculations for higher order moments as well, and we refer to it subsequently as the \textit{standard diagram}.

We see that all pairs of neighbouring many-particle states are connected by one bond $\feyn{f}$ indicating that they have to share $m-k$ single-particle states and by one bond $\feyn{h}$ in a different colour indicating $k$ shared states. A priori it is not clear how these two sets of shared single-particle states are related. However we have seen that in the limit $l\to\infty$ we only need to consider the configuration maximising the overall number of single-particle states participating in the diagram under consideration.
We can now use the fact that, by the reasons given in subsection \ref{embedded_subsection},
all single-particle states participating in the diagram are determined by
$|\mu\rangle$ and $|\rho\rangle$. Hence the number of single-particle states becomes maximal if $|\mu\rangle$ and $|\rho\rangle$ are as distinct as possible. If we now consider $|\nu\rangle$ we see that each single-particle state in $|\nu\rangle$ is included in one of the disjoint bonds $\mu\feyn{h}\nu$ and $\nu\feyn{f}\rho$, and in  one of the disjoint bonds $\mu\feyn{f}\nu$ and $\nu\feyn{h}\rho$. However if any single-particle state is included in a bond leading from $|\nu\rangle$ to $|\mu\rangle$ and a bond leading from $|\nu\rangle$ to $|\rho\rangle$ this gives rise to an overlap between $|\mu\rangle$ and $|\rho\rangle$. Hence in order to keep $|\mu\rangle$
and $|\rho\rangle$ as distinct as possible we are compelled to include as many states as possible in two bonds leading to the same node $|\mu\rangle$ or $|\rho\rangle$.
This is achieved if e.g. the smaller set of $\min(k,m-k)=:r$ single-particle states shared by $|\mu\rangle$ and $|\nu\rangle$ is included in the larger set of $\max(k,m-k)$ states shared by the same two many-particle  states. The same reasoning applies to all the bonds between neighbouring states.

In each case this leaves $s :=\max(k,m-k)-\min(k,m-k)=|m-2k|$ states  present only in the larger set but not the smaller one. We will now show that these $|m-2k|$ single-particle states
are the same for all pairs of neighbouring nodes and form the minimum possible overlap between $|\mu\rangle$ and $|\rho\rangle$.
To explain this let us first assume  that we have $m\geq2k$ and thus $r=k$, $s=m-2k$ meaning that each bond $\feyn{f}$  involves $k$ single-particle states also included in the corresponding bond $\feyn{h}$ as well as $m-2k$ further states. We now consider e.g. the $m-2k$ states included in the outer
(red) bond $\mu\feyn{f}\nu$ but not in the inner (black) bond $\mu\feyn{h}\nu$. As these states are contained in $|\nu\rangle$ our earlier reasoning implies that they have to be contained in exactly one bond of each colour attached to $|\nu\rangle$. This means that they have to be part of the  bond $\nu\feyn{f}\rho$ but they may not be part of the  bond $\nu\feyn{h}\rho$. Hence they are also part of the neighbouring set of states included in the larger bond but not the smaller one. Arguing like this for all further pairs of neighbouring states and generalizing the same reasoning to $m<2k$, we see that there is just one common set of $s=|m-2k|$ single-particle states shared by all larger bonds and omitted from all smaller bonds in the diagram.

To sum over all relevant choices of states giving non-zero contributions we have to sum over all ways to partition $l$ available single particle states into four different sets of $r$ states, one set of $s$ states, and the rest. The total number of selected single-particle states $4r+s$ equals the required value $m+2k$ only for $m\geq 2k$, and is lower for $m<2k$.
Considering the former case the number of partitions is given by the multinomial
(see (\ref{multi1}))\begin{equation}
\label{standard_unnormalized}
A_{\mu\nu\rho\sigma}A_{\sigma\mu\nu\rho}={l \choose k\;k\;k\;k\; m-2k}\sim\frac{1}{k!^4m!}{m\choose
k}{m-k\choose k}\left(\frac{l}{e}\right)^{m+2k}.
\end{equation}
For later use we mention that this multinomial can  alternatively be factorised into
${l\choose m}{m \choose k\; k}{l-m\choose k\;k}$.
Here the first factor gives the possible choices for one of the many-particle states, the second factor gives the number of ways of dividing its states into the two attached bonds $\feyn{h}$ and the set of $m-2k$ single-particle states shared by all many-particle states, and the third factor gives the number of choices for the remaining two bonds $\feyn{h}$.

After normalization we obtain (see (\ref{multi2}))
\begin{equation}
 \frac{\frac{1}{N}A_{\mu\nu\rho\sigma}A_{\sigma\mu\nu\rho}}{\left(\frac{1}{N}\mathrm{tr}(\overline{V_k^2})\right)^2}
 ~\sim\frac{{m-k\choose k}}{{m\choose k}}.\label{standard_result}
\end{equation}
which is also true even in the case $m<2k$. In this case the numerator gives zero, commensurate with the fact
 that the number of participating states is too low to give a contribution as $l\to\infty$.

\subsection{Final result and limiting cases}
\begin{figure}[h!]
\centering
\includegraphics[scale=.6]{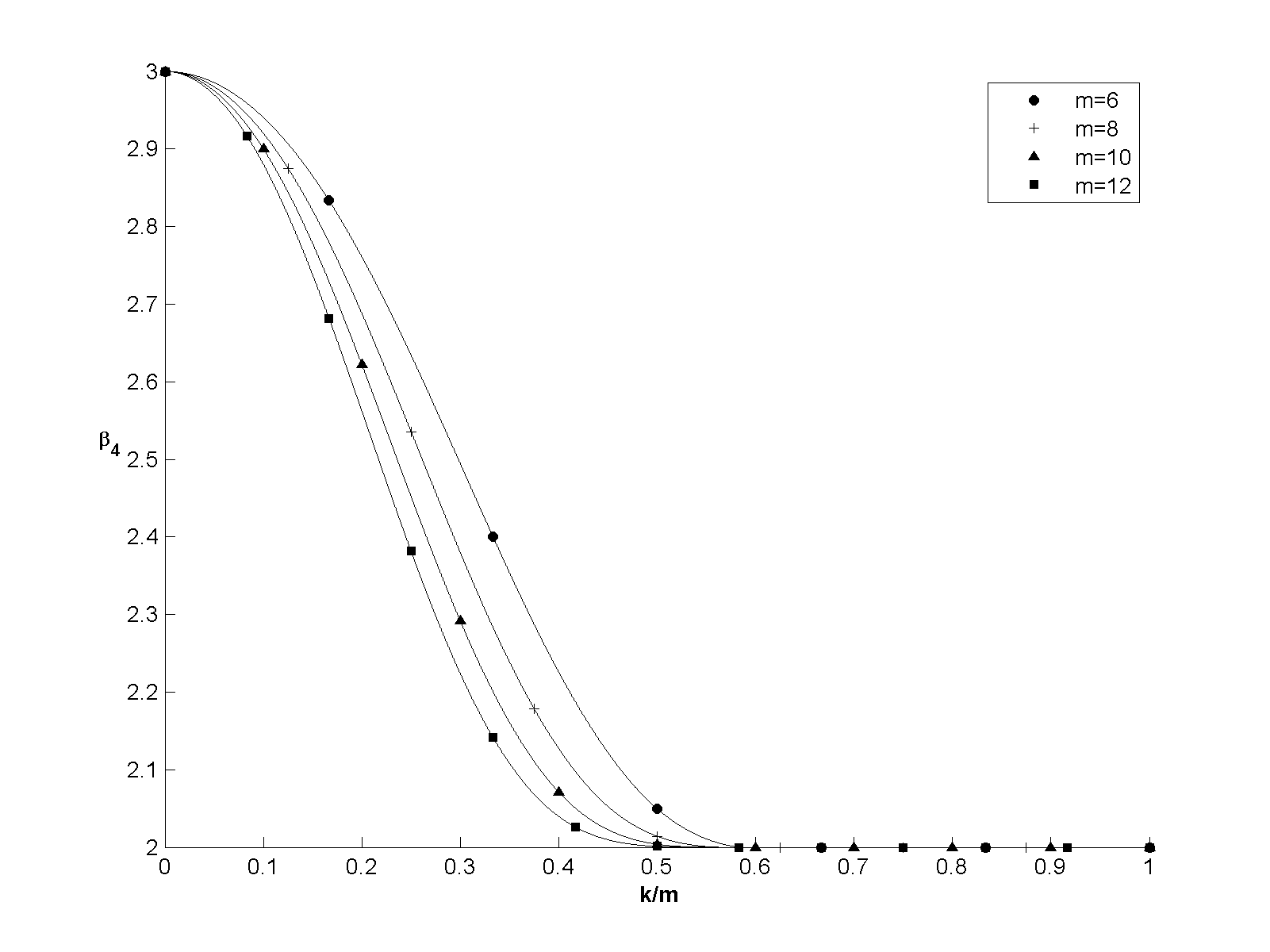}
\caption{$\kappa$ against $k/m$ for $m=6, 8, 10, 12$ showing how the kurtosis has a semi-circular value, $\kappa=2$ for $k/m > \frac{1}{2}$ after which it transitions to a Gaussian kurtosis, $\kappa = 3$ at $k=0$. Higher values of $m$ give faster convergence to the semi-circular moment.}
\label{fig:fourth_moment}
\end{figure}

Together with (\ref{reduced_example}) the overall result for the kurtosis is
\begin{align}\label{eq:arg16}
\kappa \sim 2 + \frac{{{m - k} \choose k}}{{m \choose k}}\end{align}
(depicted in Fig. \ref{fig:fourth_moment}) which corroborates the result found by Benet \textit{et. al.} using the eigenvector expansion method and for $m<2k$ agrees with what is expected using the method of supersymmetry, namely $\kappa=2$\cite{weid}.

As observed in \cite{weid} Eq. (\ref{eq:arg16})  suggests a crossover between a Gaussian level density in the limit $m\to\infty$ and a semi-circular level density for small $m$, specifically $m<2k$. Here the limit $m\to\infty$ is taken after the limit $l\to\infty$ as appropriate for the regime $k\ll m\ll l$; this setting is  referred to as the dilute limit. The normalised moments for a Gaussian level density are
$\beta_{2n}=(2n-1)!!=\frac{(2n)!}{2^nn!}$ i.e. $\kappa=\beta_4=3$, $h=\beta_6=15$,
$\tau=\beta_8=105$. For a semi-circular level density we have normalised moments given by the Catalan numbers $\beta_{2n}=\frac{1}{n+1}{2n\choose n}$ i.e. $\kappa=\beta_4=2$, $h=\beta_6=5$, $\tau=\beta_8=14$. The normalization of $\beta_{2n}$ makes sure that the expectations hold regardless of the width of the Gaussian or semi-circle.

Given ${m-k\choose k}/{m\choose k}\to 1$ for $m\to\infty$ and ${m-k\choose k}=0$ for $m<2k$ the present results, depicted in Fig. \ref{fig:fourth_moment} coincide with Gaussian and semi-circular moments respectively. This will also be the case for the sixth and eighth moments. Moreover one sees a general pattern for the reasons behind the crossover: In the limit $m\to\infty$ all diagrams give the same contribution 1 (apart from multiplicity factors). This was already shown in general in \cite{mon} without deriving the corrections for finite $m$ considered here, and it is at the heart of the so-called binary correlation approximation \cite{mon,weidreview}. As a consequence in the limit $m\to\infty$ each moment must coincide with the corresponding number of diagrams i.e. the number of pairwise contractions between $2n$ elements. As this number is given precisely by $(2n-1)!!$ one can immediately conclude that the average level density in the limit $l\to\infty$ is Gaussian.

Now for $m<2k$ our previous reasoning implies that diagrams with non-overlapping contraction lines still give contributions equal to 1. However  the ``standard diagram'' gives a vanishing contribution in this limit, and at least up to the eighth moment we will see that this applies to all  diagrams with overlapping contractions. It is natural to assume that this pattern continues for all moments.
If this is the case the $2n$-th moment for $m<2k$ must coincide with the number of non-overlapping contractions between $2n$ elements. As this number is precisely $\frac{1}{n+1}{2n\choose n}$ \cite{kreweras} one thus obtains a semi-circular average level density.

\subsection{Bosonic states}
\label{bosons}
In bosonic systems many-particle states may contain multiple copies of the same single-particle state. Retaining the same potential as before (\ref{eq:ee27}), we will now embed it instead within a \textit{bosonic} state space. This is in line with the definition of the eGUE in the fermionic case but slightly differs from the convention adopted in \cite{weidfermi}. For the case considered it is straightforward to see that in the limit $l\to\infty$ the results for all moments agree with fermionic systems. Intuitively, this happens because in this limit contributions  arise only for those choices of many-particle states that maximise the number of participating single-particle states. This means that in the bosonic case multiple occupancy of the same single-particle states is avoided and there is no difference from the fermionic case.

For a formal derivation let us consider bosonic $m$-particle states containing repeats of $z\le m$ unique single-particle states. There are ${l \choose z}$ ways to select the participating single-particle states, and ${m-1\choose z-1}$ ways to select their multiplicities to obtain altogether $z$ particles. (To understand the latter factor assume that the $m$ particles are ordered in such a way that particles in coinciding states form groups following each other. Then the first particle is definitely at the start of a group, and there are ${m-1\choose z-1}$ ways to select the starts of the remaining $z-1$ groups.) The number of available many-particle states is thus
\begin{align}\label{eq:arg18}N = \sum_{z=1}^{m}{l\choose z}{{m-1}\choose{z-1}} = {{l+m-1}\choose{m}}\end{align}
where the overall sum as well as the summand $z=m$ have the argument $m$ whereas all other summands have lower arguments. As a consequence the asymptotic form of $N$ coincides with the fermionic case where $N={l\choose m}$.

The same logic applies when evaluating the particle diagrams. In the fermionic case, the contributions to the particle diagrams with maximal argument $m+nk$ can always be factorized into a term ${l\choose {m+nk}}$ counting the number of ways in which the $m+nk$ participating states can be chosen, and further $l$-independent factors counting the number of ways in which these $m+nk$ states can be distributed among different sets while obeying the conditions implied by the diagram. In the bosonic case we instead have to  select $1\leq z\leq m+nk$ states with multiplicities summing to $m+nk$. The contribution of each diagram turns into a sum over $z$, and for each $z$ we obtain a single $l$-dependent term ${l\choose z}$ and further $l$-independent finite factors counting  the number of ways to distribute these states among sets containing $m$ particles. The latter choices have to be compatible both with the restrictions implied by the diagram and the requirement on the multiplicities. However, the only summand attaining the maximal argument will be the last, for which $z=m+nk$. For this summand all multiplicities are 1 meaning that we only have to consider bosonic states composed entirely of distinct single-particle states. Since these are mathematically equivalent to fermionic states we can conclude without any further calculations that \textit{all bosonic and fermionic moments  are equal in the limit} $l\to\infty$.

A different definition of the eGUE for bosons is considered in \cite{weidfermi} allowing for $\bbi$ and $\bj$ in (\ref{eq:egue01}) to contain repeated elements as well. In order to retain agreement with canonical RMT for $m=k$ this requires one to introduce normalisation factors depending on the number of repeated states; these factors become $1$ if all states in $\bbi$  and $\bj$ are distinct. The modified definition requires a
more significant change of our formalism that will not be pursued here; in particular the meaning of the bonds $\feyn{f}$ and $\feyn{h}$ must be adapted. However in the limit $l\to\infty$ the number of additional summands added to the $k$-body potential due to this modification is negligible compared to the number of terms without repetitions inside $\bbi$ and $\bj$. For this reason we expect that that the results obtained in this limit will also carry over to the definition of \cite{weidfermi} and again give agreement with the fermionic case. 

\section{Paths, loops and the sixth moment}
 If the fourth moment confirmed that the method of particle diagrams can greatly simplify calculations, the sixth moment illustrates the flexibility of the method as well as its ability to scale. For the sixth moment we transition from flat particle diagrams to three-dimensional graph-like diagrams. Indeed, we will begin to  use the terminology of graphs in the following calculations. In addition
to  $nodes$   (states) and {\it bonds} we will also consider $paths$ (sequences of neighbouring nodes) and  $loops$ on the particle diagrams (first state in path = last state in path), and these will become important tools for maximising the argument of attendant binomial expressions.

\begin{figure}[b]
\centering
\includegraphics[scale=.5]{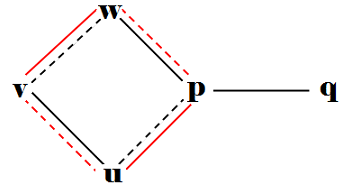}
\caption{Particle diagram for $A_{ptqq}A_{twvu}A_{upwv}$; the diagram requires
 $|p\rangle=|t\rangle$ to contribute and it is depicted only for this case.
}
\label{fig:tail}
\end{figure}

The sixth moment of the level density is given by
\begin{equation}\label{eq:s1}h = \frac{\frac{1}{N}\mathrm{tr}(\overline{V_{k}^6})}{\left(\frac{1}{N}\mathrm{tr}(\overline{V_{k}^2})\right)^3}\end{equation}
and using Wick's theorem  we obtain
\begin{align}
\mathrm{tr}(\overline{V_k^6})
&=2 \langle\mathrm{tr}
\contraction[2ex]{}{V_k}{}{V_k}
\contraction[2ex]{V_kV_k}{V_k}{}{V_k}
\contraction[2ex]{V_kV_kV_kV_k}{V_k}{}{V_k}
V_k V_k V_k V_kV_k V_k\rangle
+
3\langle\mathrm{tr}
\contraction[2ex]{}{V_k}{}{V_k}
\contraction[4ex]{V_kV_k}{V_k}{V_kV_k}{V_k}
\contraction[2ex]{V_kV_kV_k}{V_k}{}{V_k}
V_k V_k V_k V_kV_k V_k\rangle
+
6\langle\mathrm{tr}
\contraction[2ex]{}{V_k}{}{V_k}
\contraction[2ex]{V_kV_k}{V_k}{V_k}{V_k}
\contraction[4ex]{V_kV_kV_k}{V_k}{V_k}{V_k}
V_k V_k V_k V_kV_k V_k\rangle\nonumber\\
&+
3\langle\mathrm{tr}
\contraction[2ex]{}{V_k}{V_k}{V_k}
\contraction[2ex]{V_k V_k V_k}{V_k}{V_k}{V_k}
\contraction[4ex]{V_k}{V_k}{V_k V_k}{V_k}
V_k V_k V_k V_kV_k V_k\rangle
+\langle
\mathrm{tr}
\contraction[2ex]{}{V_k}{V_kV_k}{V_k}
\contraction[4ex]{V_k}{V_k}{V_kV_k}{V_k}
\contraction[6ex]{V_kV_k}{V_k}{V_kV_k}{V_k}
V_k V_k V_k V_kV_k V_k\rangle.
\end{align}
Here the prefactors indicate the number of equivalent diagrams that can be obtained by cyclic permutation of the trace. Written in terms of $A_{\mu\nu\rho\sigma}$ the summands are as follows:
\begin{align}\label{eq:s3} \mathrm{tr}(\overline{V_{k}^6}) & = 2A_{ptqq}A_{tvuu}A_{vpww} + 3A_{ptqq} A_{tpwu}A_{uwvv}
+ 6A_{ptqq}A_{twvu}A_{upwv} \nonumber\\&+ 3A_{putq}A_{qwvt}A_{upwv}  + A_{pvuq}A_{qwvt}A_{tpwu}.\end{align}
The first three terms involve contractions between neighbouring $V_k$ and hence have coinciding subscripts in some of the factors. These terms can be reduced to lower-order contributions using the results in subsection \ref{identical}.
Taking into account the normalization factors we obtain
\begin{align}
\label{trV61}
2
\frac{\frac{1}{N}A_{ptqq}A_{tvuu}A_{vpww}}{\left(\frac{1}{N}\mathrm{tr}\overline{V_k^2}\right)^3} &=2\frac{\frac{1}{N}A_{pvuu}A_{vpww}}{\left(\frac{1}{N}\mathrm{tr}\overline{V_k^2}\right)^2}
=2\\
3
\frac{\frac{1}{N}A_{ptqq}A_{tpwu}A_{uwvv} }{\left(\frac{1}{N}\mathrm{tr}\overline{V_k^2}\right)^3} &=3\frac{\frac{1}{N}A_{ppwuA_{uwvv}}}{\left(\frac{1}{N}\mathrm{tr}\overline{V_k^2}\right)^2}
=3\\
\label{trV63}
6
\frac{\frac{1}{N}A_{ptqq}A_{twvu}A_{upwv} }{\left(\frac{1}{N}\mathrm{tr}\overline{V_k^2}\right)^3}
&=6
\frac{\frac{1}{N}A_{pwvu}A_{upwv} }{\left(\frac{1}{N}\mathrm{tr}\overline{V_k^2}\right)^2}\sim 6\frac{{m-k \choose k}}{{m \choose k}}.
\end{align}
Here we used the results for the fourth moment contribution (\ref{reduced_example})
in the first two lines, and (\ref{standard_result}) in the third line.
In the first two lines  the nontrivial factors in the numerator and denominator compensate exactly and the results are integers, in line with the general behaviour if contraction lines do not intersect. The particle diagram responsible for the third line is displayed in Fig. \ref{fig:tail}; it is  identical to the ``standard diagram'' except for the addition of a tail ${p}\feyn{f}{q}$.

\subsection{The role of closed loops in particle diagrams}
\label{loops}
This leaves only two diagrams that cannot be reduced to diagrams already contributing to the fourth moment.
For these diagrams, as well as more complicated particle diagrams to follow, the task of identifying the
relevant overlaps between states becomes more and more involved, and it is desirable to have a more algorithmic approach. This can be realised by considering the particle diagram as a graph, and studying closed loops on this graph.

To understand the relevance of closed loops, consider a single-particle state $i$ that belongs to a  many-particle state $|p\rangle$ appearing in a given diagram. This $|p\rangle$ has to appear twice as a subscript in the product of $A$'s.   If it appears twice as a subscript of the same $A$ we can apply the procedure of subsection \ref{identical} to simplify the diagram and relate it to a contribution to a lower moment. Hence we can restrict ourselves to the case that the two subscripts belong to different factors.

 In the particle diagram each of these factors $A$ is represented by a square as in Fig. \ref{fig:square}, and in this square one bond $\feyn{f}$ and one bond $\feyn{h}$ are attached to $|p\rangle$. The single-particle state $i$ is included in exactly one of these bonds. As this applies to both squares the state $i$ is thus contained in a path that so far contains $|p\rangle$ and two attached bonds belonging to different $A$'s. However now our reasoning can be continued as $i$ must also be contained in the many-particle states at the other end of each of the two bonds and then a further bond also attached to these states. This continues until all bonds included in the path form a closed loop. All loops in the particle diagram are permissible on condition that subsequent bonds along the loop belong to different $A$'s (depicted in different colours in our figures). The shortest loops consist of two bonds that connect the same two nodes but belong to different $A$'s. The longest loops visit all nodes contained in the graph.

By the above argument each single-particle state contributing to one of the many-particle states belongs to a closed loop. In fact we can assume that it belongs to precisely one loop. It clearly cannot belong to two different loops that intersect in some nodes -- this would mean that for some states more than two of the attached bonds share single-particle states, which for fermionic states is ruled out. It would in principle be possible to have the same state included in two loops that visit disjoint sets of nodes. However this would decrease the number of participating single-particle states and hence the argument of the diagram. Hence we can ignore this possibility in the limit $l\to\infty$.

Now the most general method for evaluating particle diagrams is the following: First, determine all closed loops subject to the constraint above. Our notation for a loop starting from the node $p_1$ and proceeding through $p_2,\ldots,p_M$ before returning to $p_1$ will be $\overrightarrow{p_1p_2\ldots p_M}$; as the loops are only used to describe overlaps between the bonds, loops with reverted ordering are considered identical. We introduce variables $n_j$ for the number of single-particle states participating in each loop.
The sum $\sum_j n_j$ gives the argument. Then we incorporate the constraint that each bond $\feyn{h}$ contains $k$ states and each bond $\feyn{f}$ contains $m-k$ states. This constraint fixes the sum of the $n_j$'s belonging to all loops that contain a given bond, and thus leads to a system of linear equations for the $n_j$'s. If we solve this system of equations it yields all $n_j$'s, and therefore also the argument, as a linear function of a reduced number of parameters.
These parameters  have to be chosen in such a way that the argument is maximal and this optimal choice will often involve vanishing $n_j$'s i.e. not all permitted loops have to contribute.
If one obtains  a unique choice for all $n_j$'s we have to consider all ways of selecting from the $l$ single particle states the required sets of $n_j$ states. The contribution of the diagram (prior to normalization) is then given by
\begin{equation}
\label{multi}
{l\choose n_1\;n_2\;n_3\;\ldots}.
\end{equation}
If  free parameters remain even after maximising the argument we obtain a sum over multinomials representing the remaining choices of $n_j$. However in several of the cases to be considered there are  shortcuts to the solution so  that some of these steps are not needed.

\subsection{Diagram $A_{putq}A_{qwvt}A_{upwv}$}
\label{prism}
\begin{figure}[b]
\centering
\includegraphics[scale=.5]{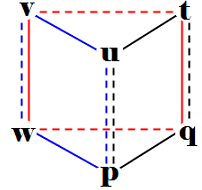}
\caption{Particle diagram for $A_{putq}A_{qwvt}A_{upwv}$.}
\label{fig:prism}
\end{figure}

\begin{figure}[ht!]
\minipage{0.25\textwidth}
\centering
(a)\includegraphics[scale=0.25]{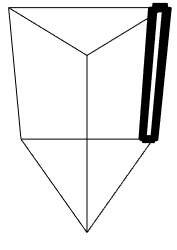}
\endminipage\hfill
\minipage{0.25\textwidth}
\centering
(b)\includegraphics[scale=.25]{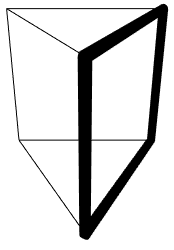}
\endminipage\hfill
\minipage{0.25\textwidth}
\centering
(c)\includegraphics[scale=0.25]{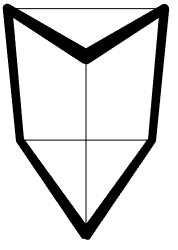}
\endminipage\hfill
\minipage{0.25\textwidth}
\centering
(d)\includegraphics[scale=.25]{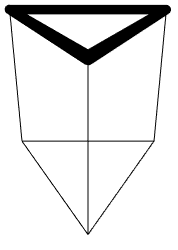}
\endminipage\hfill
\caption{Illustrations of the loops
(a) $\protect\arr{tq}$, (b) $\protect\arr{utqp}$, (c) $\protect\arr{vutqpw}$, (d) $\protect\arr{tvu}$.
By maximising the argument we find that the loops (a) and (d) contain $k$ single-particle states. The loops (b) and (c) are included in Eq. (\ref{sysred}), and the numbers of states in these loops are not determined uniquely.}
\label{fig:loopsprism}
\end{figure}

The particle diagram $A_{putq}A_{qwvt}A_{upwv}$ can still be evaluated without using closed loops (see \cite{small}) but  we want to use it here to illustrate the path summation method introduced
above. The diagram has the structure of a triangular prism as shown in Fig. \ref{fig:prism}.
In the figure the faces on the sides correspond to the different factors $A$ and are thus depicted in different colours. First we need to consider all loops permitted by the diagram. The simplest loops just move up and down using the two bonds on the vertical sides of the prism.
These loops (see e.g. Fig. \ref{fig:loopsprism}a) are denoted by
\begin{equation}
\overrightarrow{tq},\overrightarrow{vw},\overrightarrow{up}.
\end{equation}
Next there are loops that move for a while along the bonds on the top face of the prism but then then move down through one of the bonds on the side. Interestingly the motion at the bottom of the prism is determined by the motion on the top. For example a loop containing the bond $u\feyn{f}t$ can reach the node $|q\rangle$ in the following step only through the bond $t\feyn{f}q$ as the bond $t\feyn{h}q$ belongs to the same factor $A$ as the preceding one. Then the only possible choice for the next bond is $q\feyn{f}p$.
 Analogous reasoning holds for all sides, meaning that if a loop travelling along the top of the diagram moves down it is afterwards ``reflected'' and follows the image of the previous bonds on the bottom face. This leads to a number of further loops   given by (see e.g. Fig. \ref{fig:loopsprism}b and
\ref{fig:loopsprism}c)
\begin{align}
&\overrightarrow{tv\bo{wq}}, \overrightarrow{vu\bo{pw}}, \overrightarrow{ut\bo{qp}}
,\nonumber\\ &\overrightarrow{vut\bo{qpw}}, \overrightarrow{utv\bo{wqp}}, \overrightarrow{uvt\bo{qwp}}, \overrightarrow{tvu},\overrightarrow{pwq};
\end{align}
here the first half of the indices describes the motion on the top face and
the second half describes the motion in the opposite direction at the bottom.

Finally there are loops travelling only on the top face ($\overrightarrow{tvu}$, see Fig. \ref{fig:loopsprism}d) or only on the bottom face ($\overrightarrow{qwp}$).
The constraints for the division of states among  loops can  be solved consistently only if the numbers of states in the two final loops coincide. This follows from the fact that  all other loops visit the top and bottom face symmetrically, and all other bonds have the same symmetry. The number of states included in each of the final two loops will be denoted by $n_{\rm top}$.

Now we have to incorporate the constraint that every bond $\feyn{f}$ contains $m-k$ states and every bond $\feyn{h}$ contains $k$ states. This fixes the sum of the numbers of states included in the loops traversing each bond. For the bonds on the top face of the prism this leads to the three equations
\begin{align}
\label{sys1}
n_{tv\bo{wq}}+n_{utv\bo{wqp}}+n_{tvu\bo{pwq}}+n_{\rm top}&=k\nonumber\\
n_{vu\bo{pw}}+n_{tvu\bo{pwq}}+n_{vut\bo{qpw}}+n_{\rm top}&=m-k\nonumber\\
n_{ut\bo{qp}}+n_{vut\bo{qpw}}+n_{utv\bo{wqp}}+n_{\rm top}&=m-k\;.
\end{align}
The equations arising from the bonds on the bottom are identical to these and hence do not give further information. The equations arising from the six bonds on the sides are
\begin{align}
\label{sys2}
n_{t\bo{q}}+n_{tv\bo{wq}}+n_{tvu\bo{pwq}}&=k\nonumber\\
n_{v\bo{w}}+n_{vu\bo{pw}}+n_{vut\bo{qpw}}&=m-k\nonumber\\
n_{u\bo{p}}+n_{ut\bo{qp}}+n_{utv\bo{wqp}}&=k\nonumber\\
n_{t\bo{q}}+n_{ut\bo{qp}}+n_{vut\bo{qpw}}&=m-k\nonumber\\
n_{v\bo{w}}+n_{tv\bo{wq}}+n_{utv\bo{wqp}}&=k\nonumber\\
n_{u\bo{p}}+n_{vu\bo{pw}}+n_{tvu\bo{pwq}}&=k\;.
\end{align}
The argument is simply the total number of states participating in all loops, $\sum n_j$, including $n_{\rm top}$ twice as there are two loops with this number of states. Solving the system of equations provided by (\ref{sys1}) and (\ref{sys2}) allows one to express six of the loop numbers as linear functions of the others. Summation then leads to the argument as a function of the remaining variables. If  we include $n_{\rm top}$ among the variables that are not eliminated the result for the argument is $m+k+2n_{\rm top}$. Hence we can maximise the argument by choosing
\begin{align}
n_{\rm top}=k
\end{align}
 which is the largest possible value for $n_{\rm top}$ as the corresponding loop contains the bond $t\feyn{h}v$ with $k$ states.
This result implies that all other loops containing the bond $t\feyn{h}v$ (and appearing in the first line of (\ref{sys1})) must be empty, i.e.,
\begin{equation}
n_{tv\bo{wq}}=n_{utv\bo{wqp}}=n_{tvu\bo{pwq}}=n_{uvt\bo{qwp}}=0.
\end{equation}
With $t\feyn{h}v$ (and analogously $q\feyn{h}w$) accounted for, the only way to incorporate the bond $t\feyn{h}q$ in a loop is to combine it with $t\feyn{f}q$ to form $\overrightarrow{tq}$. This loop must therefore contain all $k$ states of $t\feyn{h}q$ leading to
\begin{equation}
n_{t\bo{q}}=k.
\end{equation}
An analogous argument gives
\begin{equation}
n_{v\bo{w}}=k.
\end{equation}
If we insert these results into the system of equations above it reduces to
\begin{align}
\label{sysred}
n_{u\bo{p}}+n_{ut\bo{qp}}+n_{vu\bo{pw}}+n_{vut\bo{qpw}}&=m-k\nonumber\\
n_{u\bo{p}}+n_{vu\bo{pw}}&=k\nonumber\\
n_{u\bo{p}}+n_{ut\bo{qp}}&=k
\end{align}
where the final two equations fix the number of states associated to the
two bonds $u\feyn{h}p$.

Having three equations for four unknowns we could now parametrise the solutions by  one variable,  express the result as a sum over multinomials depending on this variable, and then simplify the sum. However the form of the equation allows us to evaluate the diagram more quickly.

We have to consider all ways to select from altogether $l$ single-particle states four loops with $k$ states each (along the sides of the top and bottom face, and the two loops $\overrightarrow{tq}$ and $\overrightarrow{vw}$)
and $m-k$ states that according to the first line of (\ref{sysred}) have to be distributed among the loops  $\arr{up}$, $\arr{utqp}$, $\arr{vupw}$, and $\arr{vutqpw}$. This leads to a factor ${l\choose k\;k\;k\;k\;m-k}$.
Then, given the second line of (\ref{sysred}) we have to select $k$ of the $m-k$ states in $\arr{up}$, $\arr{utqp}$, $\arr{vupw}$, and $\arr{vutqpw}$ to be included in $\arr{up}$ and $\arr{vupw}$. There are ${m-k\choose k}$ ways to do this. Due to the third line of (\ref{sysred}) we independently have to select $k$ of the same $m-k$ states to be included in $\arr{up}$ and $\arr{utqp}$, giving rise to a second factor ${m-k\choose k}$. These two selections uniquely determine how the final $m-k$ single-particle
states are distributed among the four loops in question, i.e. states selected twice are included in $\arr{up}$ and states omitted from both selections are included in $\arr{vutqpw}$.
Altogether we thus obtain
\begin{equation}
A_{putq}A_{qwvt}A_{upwv}\sim{l\choose k\;k\;k\;k\;m-k}{m-k\choose k}^2
\end{equation}
and, after normalization (see (\ref{multi2})),
\begin{equation}\label{trV64}
\frac{\frac{1}{N}A_{putq}A_{qwvt}A_{upwv}}{\left(\frac{1}{N}\mathrm{tr}(\overline V^2)\right)^3}\sim\frac{{m-k\choose k}^2}{{m\choose k}^2}.
\end{equation}

\subsection{Diagram $A_{pvuq}A_{qwvt}A_{tpwu}$}
\label{diamond}
\begin{figure}
\centering
\includegraphics[scale=0.5]{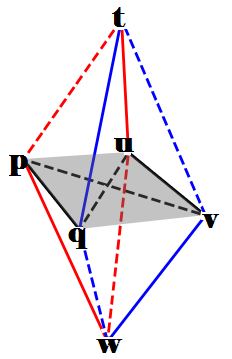}
\caption{The particle diagram for the term $A_{pvuq}A_{qwvt}A_{tpwu}$ which takes the form of a regular octahedron, or two square pyramids with a shared base determined by the plane on which the sub-diagram for $A_{pvuq}$ is illustrated. The states $|t\rangle$ and $|w\rangle$ together determining the states $|v\rangle$ and $|q\rangle$ through the bonds defined by $A_{qwvt}$ just as they determine the states $|u\rangle$ and $|p\rangle$ through the bonds defined by $A_{tpwu}$. 
}
\label{fig:diamond}
\end{figure}

The only remaining particle diagram contributing to the sixth moment is $A_{pvuq}A_{qwvt}A_{tpwu}$, depicted in Fig. \ref{fig:diamond}. In this case there is a convenient shortcut that will allow us to find the optimal numbers of states participating in the loops.
The key to this shortcut is that each participating single-particle state must be included in either $|w\rangle$ or $|t\rangle$ or both.
To understand this we recall that for each quadrangle of many-particle states representing one of the factors $A$, the union of two diametrically opposed states coincides with the union of the two other states.
Applying this to the blue quadrangles we see that  $|q\rangle$
and $|v\rangle$  contain only states included in the union of $|w\rangle$ and $|t\rangle$. Considering the red quadrangle gives the same result for $|p\rangle$ and $|u\rangle$.
To maximise the number of participating single-particle states we thus have to make $|t\rangle$ and $|w\rangle$ as distinct as possible.
This means that we should include as many single-particle states as possible
in loops that contain only one of these nodes.
There are four loops containing only $|w\rangle$,
\begin{equation}
\overrightarrow{wvp}, \overrightarrow{wvu}, \overrightarrow{wqp}, \overrightarrow{wqu},
\end{equation}
and four loops containing only $|t\rangle$,
\begin{equation}
\overrightarrow{tuq},\overrightarrow{tuv},\overrightarrow{tpq}, \overrightarrow{tpv}.
\end{equation}
Assuming that $m>3k$, the number of states in these loops can be maximised if we include the maximum permissible number of $k$ states in the first three loops in each group, leaving the final loops $\overrightarrow{wqu}$ and $\overrightarrow{tpv}$ empty.
This choice is optimal because  the omitted loops each involve only bonds $\feyn{h}$ that contain $k$  states that are each shared with one other loop from the same group. Hence including one state in, say, $\overrightarrow{wqu}$
would reduce the maximal number of states that can be included in $\overrightarrow{wvp}$, $\overrightarrow{wvu}$, and $\overrightarrow{wqp}$ by one each.
Choosing $k$ states for the first three loops in each of the two groups uniquely determines all bonds $\feyn{h}$ in the diagram as each of them contains $k$ states and is included in precisely one of these loops. All bonds $\feyn{f}$ contain $m-k$ single-particle states and participate in two of the loops with $k$ states each, leaving $m-3k$ states undetermined. As the bonds $\feyn{f}$ together form one single loop $\overrightarrow{wpqtuv}$ the number of states included in this loop must also be $m-3k$.

To evaluate the present diagram we thus have to consider all ways to select from $l$ states  six loops of $k$ states and one loop of $m-3k$ states implying
\begin{equation}\label{eq:sixth_sr}A_{pvuq}A_{qwvt}A_{tpwu} \sim {{l}\choose{k~k~k~k~k~k~m-3k}}.\end{equation}
Using (\ref{multi2}) this leads to
\begin{equation}
\label{trV65}\frac{{\frac{1}{N}}A_{pvuq}A_{qwvt}A_{tpwu}}{\left(\frac{1}{N}\mathrm{tr}(\overline{V_{k}^2})\right)^3}
\sim\ \frac{{{m-k}\choose{k}}{{m-2k}\choose{k}}}{{m\choose k}^2}.\end{equation}
We still have to check our result also holds in the case $m<3k$. We use  that the argument of $A_{pvuq}A_{qwvt}A_{tpwu}$ is the overall number of states in $|w\rangle$ and $|t\rangle$, i.e., it is smaller or equal to $2m$. For $m<3k$ it is thus smaller than $m+3k$ which according to subsection \ref{arguments} is the required argument for the diagram to contribute in the limit $l\to\infty$.
Hence the result  vanishes, which is in line with (\ref{trV65})
due to ${m-2k\choose k}=0$ for the case at hand.

\subsection{Final result}
\begin{figure}[H]
\centering
\includegraphics[scale=.6]{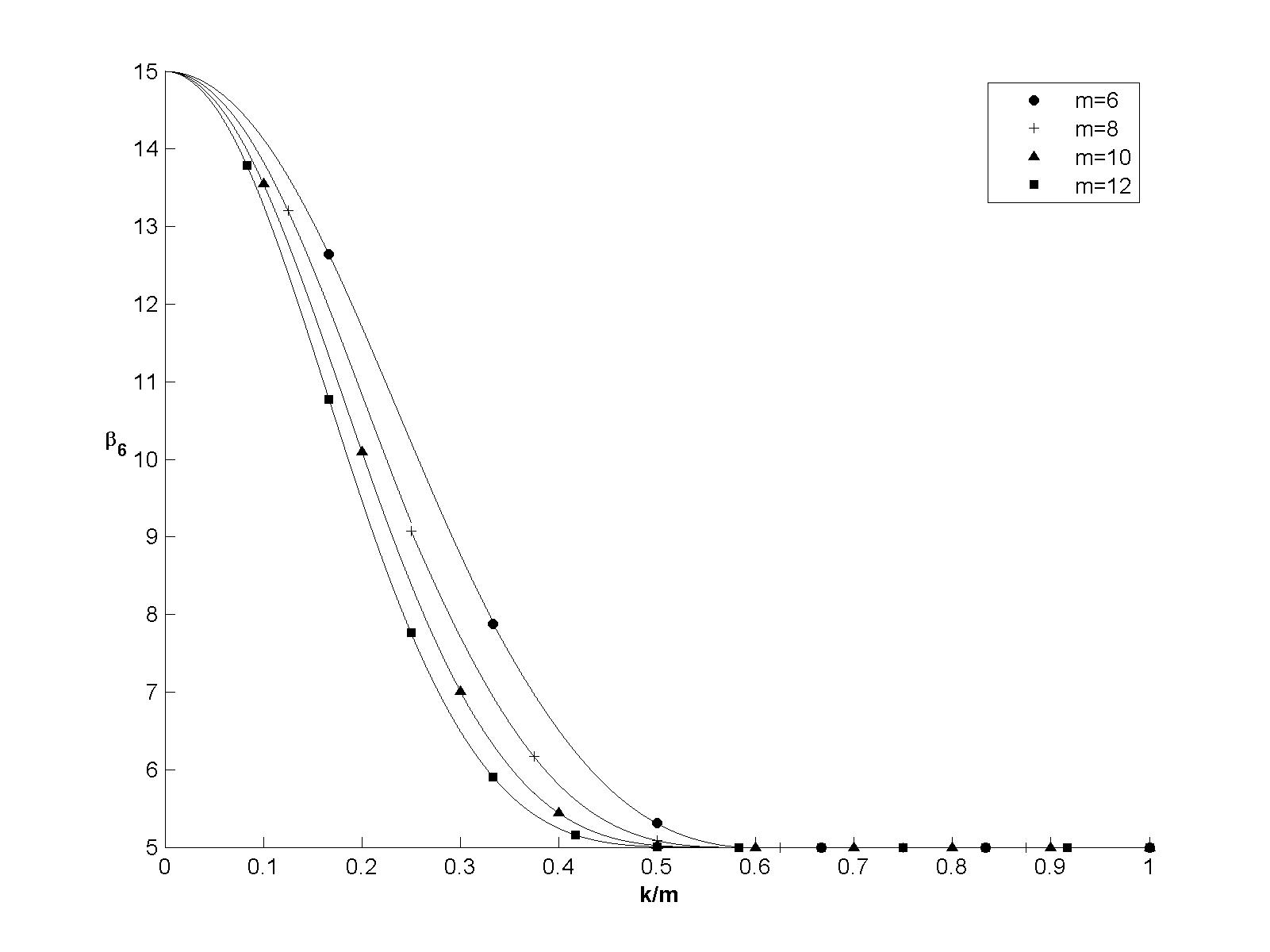}
\caption{$h$ against $k/m$ for $m=6, 8, 10, 12$. Once again we have a domain $k/m > \frac{1}{2}$ for which the sixth moment takes the semi-circular value $h=5$ and a transition thereafter towards a gaussian moment, $h=15$ at $k=0$. Higher values of $m$ give faster convergence to the semi-circular moment.}
\label{fig:sixth_moment}
\end{figure}

Together with the previous results we thus obtain the sixth moment depicted
in Fig. \ref{fig:sixth_moment}, \begin{equation}\label{eq:s8} h \sim 5 + 6\frac{{{m-k}\choose{k}}}{{m\choose k}} + 3\frac{{{m-k}\choose{k}}^2}{{m\choose
k}^2} + \frac{{{m-k}\choose{k}}{{m-2k}\choose{k}}}{{m\choose
k}^2}.\end{equation}
Again we observe a crossover from the semi-circular moment 5 (arising from
non-crossing contractions) to the Gaussian moment 15.
\section{The eighth moment}
The normalised eighth moment is given by
\begin{equation}\tau = \frac{\frac{1}{N}\mathrm{tr}(\overline{V_{k}^8})}{\left(\frac{1}{N}\mathrm{tr}(\overline{V_{k}^2})\right)^4}.\end{equation}
 It will exhibit the same features as the lower moments, in particular a transition  from a semi-circular moment $\tau = 14$ to a Gaussian moment $\tau=105$, through terms that start to contribute immediately after $m=2k,
3k, 4k$. The final one of these terms arises from a single contribution to $\mathrm{tr}(\overline{V_k^8})$ taking the form
${{l}\choose{k~k~k~k~k~k~k~k~m-4k}}$
analogous to (\ref{standard_unnormalized}) and (\ref{eq:sixth_sr}).
Again many contributions can be evaluated  by reduction to diagrams arising
for lower moments. For the remaining ones we will use the path summation
method, sometimes accompanied by helpful shortcuts. Specifically, Wick's theorem gives (dropping the indices $k$)
\begin{align}
\label{a}
\mathrm{tr}(\overline{\Vk^8})&=
2\langle\mathrm{tr}
\contraction[2ex]{}{\Vk}{}{\Vk}
\contraction[2ex]{\Vk\Vk}{\Vk}{}{\Vk}
\contraction[2ex]{\Vk\Vk\Vk\Vk}{\Vk}{}{\Vk}
\contraction[2ex]{\Vk\Vk\Vk\Vk\Vk\Vk}{\Vk}{}{\Vk}
\Vk \Vk \Vk \Vk\Vk \Vk\Vk\Vk\rangle
+
8\langle\mathrm{tr}
\contraction[2ex]{}{\Vk}{}{\Vk}
\contraction[2ex]{\Vk\Vk}{\Vk}{}{\Vk}
\contraction[4ex]{\Vk\Vk\Vk\Vk}{\Vk}{\Vk\Vk}{\Vk}
\contraction[2ex]{\Vk\Vk\Vk\Vk\Vk}{\Vk}{}{\Vk}
\Vk \Vk \Vk \Vk\Vk \Vk\Vk\Vk\rangle
+4\langle\mathrm{tr}
\contraction[2ex]{\Vk\Vk\Vk}{\Vk}{}{\Vk}
\contraction[4ex]{\Vk\Vk}{\Vk}{\Vk\Vk}{\Vk}
\contraction[6ex]{\Vk}{\Vk}{\Vk\Vk\Vk\Vk}{\Vk}
\contraction[8ex]{}{\Vk}{\Vk\Vk\Vk\Vk\Vk\Vk}{\Vk}
\Vk \Vk \Vk \Vk\Vk \Vk\Vk\Vk\rangle
\\
&+8\langle\mathrm{tr}
\contraction[2ex]{}{\Vk}{}{\Vk}
\contraction[2ex]{\Vk\Vk}{\Vk}{}{\Vk}
\contraction[4ex]{\Vk\Vk\Vk\Vk\Vk}{\Vk}{\Vk}{\Vk}
\contraction[2ex]{\Vk\Vk\Vk\Vk}{\Vk}{\Vk}{\Vk}
\Vk \Vk \Vk \Vk\Vk \Vk\Vk\Vk\rangle
+
8\langle\mathrm{tr}
\contraction[4ex]{}{\Vk}{\Vk\Vk}{\Vk}
\contraction[2ex]{\Vk}{\Vk}{}{\Vk}
\contraction[4ex]{\Vk\Vk\Vk\Vk\Vk}{\Vk}{\Vk}{\Vk}
\contraction[2ex]{\Vk\Vk\Vk\Vk}{\Vk}{\Vk}{\Vk}
\Vk \Vk \Vk \Vk\Vk \Vk\Vk\Vk\rangle
+8\langle\mathrm{tr}
\contraction[2ex]{}{\Vk}{}{\Vk}
\contraction[2ex]{\Vk\Vk\Vk}{\Vk}{}{\Vk}
\contraction[6ex]{\Vk\Vk\Vk\Vk\Vk}{\Vk}{\Vk}{\Vk}
\contraction[4ex]{\Vk\Vk}{\Vk}{\Vk\Vk\Vk}{\Vk}
\Vk \Vk \Vk \Vk\Vk \Vk\Vk\Vk\rangle\nonumber\\
&+4\langle\mathrm{tr}
\contraction[2ex]{}{\Vk}{}{\Vk}
\contraction[2ex]{\Vk\Vk\Vk\Vk}{\Vk}{}{\Vk}
\contraction[6ex]{\Vk\Vk\Vk}{\Vk}{\Vk\Vk\Vk}{\Vk}
\contraction[4ex]{\Vk\Vk}{\Vk}{\Vk\Vk\Vk}{\Vk}
\Vk \Vk \Vk \Vk\Vk \Vk\Vk\Vk\rangle\\
&\label{c}+8\langle\mathrm{tr}
\contraction[2ex]{}{\Vk}{}{\Vk}
\contraction[2ex]{\Vk\Vk}{\Vk}{\Vk\Vk}{\Vk}
\contraction[4ex]{\Vk\Vk\Vk}{\Vk}{\Vk\Vk}{\Vk}
\contraction[6ex]{\Vk\Vk\Vk\Vk}{\Vk}{\Vk\Vk}{\Vk}
\Vk \Vk \Vk \Vk\Vk \Vk\Vk\Vk\rangle\\
&\label{d}+8\langle\mathrm{tr}
\contraction[2ex]{}{\Vk}{}{\Vk}
\contraction[2ex]{\Vk\Vk}{\Vk}{\Vk}{\Vk}
\contraction[2ex]{\Vk\Vk\Vk \Vk \Vk}{\Vk}{\Vk}{\Vk}
\contraction[4ex]{\Vk\Vk\Vk}{\Vk}{\Vk \Vk}{\Vk}
\Vk \Vk \Vk \Vk\Vk \Vk\Vk\Vk\rangle
+16\langle\mathrm{tr}
\contraction[2ex]{\Vk}{\Vk}{}{\Vk}
\contraction[4ex]{}{\Vk}{\Vk\Vk\Vk}{\Vk}
\contraction[6ex]{\Vk\Vk\Vk}{\Vk}{\Vk\Vk}{\Vk}
\contraction[4ex]{\Vk\Vk\Vk\Vk\Vk}{\Vk}{\Vk}{\Vk}
\Vk \Vk \Vk \Vk\Vk \Vk\Vk\Vk\rangle\\
&\label{e}+4\langle\mathrm{tr}
\contraction[2ex]{}{\Vk}{\Vk}{\Vk}
\contraction[4ex]{\Vk}{\Vk}{\Vk}{\Vk}
\contraction[2ex]{\Vk \Vk \Vk \Vk}{\Vk}{\Vk}{\Vk}
\contraction[4ex]{\Vk \Vk \Vk \Vk\Vk}{\Vk}{\Vk}{\Vk}
\Vk \Vk \Vk \Vk\Vk \Vk\Vk\Vk\rangle
\\
&+\label{f}
4\langle\mathrm{tr}
\contraction[8ex]{}{\Vk}{\Vk\Vk\Vk}{\Vk}
\contraction[2ex]{\Vk\Vk\Vk}{\Vk}{\Vk}{\Vk}
\contraction[4ex]{\Vk\Vk}{\Vk}{\Vk\Vk\Vk}{\Vk}
\contraction[6ex]{\Vk}{\Vk}{\Vk\Vk\Vk\Vk\Vk}{\Vk}
\Vk \Vk \Vk \Vk\Vk \Vk\Vk\Vk\rangle
+2\langle\mathrm{tr}
\contraction[4ex]{}{\Vk}{\Vk\Vk\Vk\Vk}{\Vk}
\contraction[2ex]{\Vk}{\Vk}{\Vk\Vk}{\Vk}
\contraction[8ex]{\Vk\Vk}{\Vk}{\Vk\Vk\Vk\Vk}{\Vk}
\contraction[6ex]{\Vk\Vk\Vk}{\Vk}{\Vk\Vk}{\Vk}
\Vk \Vk \Vk \Vk\Vk \Vk\Vk\Vk\rangle
+
4\langle\mathrm{tr}\contraction[4ex]{}{\Vk}{\Vk\Vk\Vk\Vk}{\Vk}
\contraction[2ex]{\Vk}{\Vk}{\Vk\Vk}{\Vk}
\contraction[8ex]{\Vk\Vk\Vk}{\Vk}{\Vk\Vk\Vk}{\Vk}
\contraction[6ex]{\Vk\Vk}{\Vk}{\Vk\Vk\Vk}{\Vk}
\Vk \Vk \Vk \Vk\Vk \Vk\Vk\Vk\rangle\nonumber\\
&+
8\langle\mathrm{tr}
\contraction[2ex]{}{\Vk}{\Vk}{\Vk}
\contraction[4ex]{\Vk}{\Vk}{\Vk\Vk}{\Vk}
\contraction[2ex]{\Vk\Vk\Vk}{\Vk}{\Vk\Vk}{\Vk}
\contraction[4ex]{\Vk\Vk\Vk\Vk\Vk}{\Vk}{\Vk}{\Vk}
\Vk \Vk \Vk \Vk\Vk \Vk\Vk\Vk\rangle
+
8\langle\mathrm{tr}
\contraction[4ex]{}{\Vk}{\Vk}{\Vk}
\contraction[2ex]{\Vk}{\Vk}{\Vk\Vk\Vk}{\Vk}
\contraction[4ex]{\Vk\Vk\Vk}{\Vk}{\Vk\Vk}{\Vk}
\contraction[6ex]{\Vk\Vk\Vk\Vk}{\Vk}{\Vk\Vk}{\Vk}
\Vk \Vk \Vk \Vk\Vk \Vk\Vk\Vk\rangle
+\langle\mathrm{tr}
\contraction[2ex]{}{\Vk}{\Vk\Vk\Vk}{\Vk}
\contraction[4ex]{\Vk}{\Vk}{\Vk\Vk\Vk}{\Vk}
\contraction[6ex]{\Vk\Vk}{\Vk}{\Vk\Vk\Vk}{\Vk}
\contraction[8ex]{\Vk\Vk\Vk}{\Vk}{\Vk\Vk\Vk}{\Vk}
\Vk \Vk \Vk \Vk\Vk \Vk\Vk\Vk\rangle
\end{align}
where the prefactors indicate the number of equivalent contributions that can be obtained by cyclic permutation of the trace; the prefactor 16 in the second term of (\ref{d}) also incorporates equivalent contributions obtained by reverting the order of $V$'s. Reassuringly the prefactors sum to $105=(8-1)!!$, the overall number of possible contractions between eight elements.

All terms from (\ref{a}) to (\ref{d}) involve contraction between neighbouring $\Vk$'s. As discussed in subsection \ref{identical} they can thus be reduced
to simpler diagrams contributing to lower moments.
As the contraction lines of the diagrams in (\ref{a}) do not intersect, all contraction lines can be removed in this way and the contribution can be reduced to 1 (times the multiplicity factor).
The diagrams in each of the groups (\ref{a}) to (\ref{d})  reduce to the same simplified diagram and hence give identical contributions. We only briefly display the calculations for the first diagram in each of the groups (\ref{a}) to (\ref{d})
\begin{align}
\frac{\frac{1}{N}\langle\mathrm{tr}
\contraction[2ex]{}{\Vk}{}{\Vk}
\contraction[2ex]{\Vk\Vk}{\Vk}{}{\Vk}
\contraction[2ex]{\Vk\Vk\Vk\Vk}{\Vk}{}{\Vk}
\contraction[2ex]{\Vk\Vk\Vk\Vk\Vk\Vk}{\Vk}{}{\Vk}
\Vk \Vk \Vk \Vk\Vk \Vk\Vk\Vk\rangle}{\left(\frac{1}{N}\mathrm{tr}(\overline {V^2})\right)^4}
&=\frac{\frac{1}{N}A_{ptqq}A_{tvuu}A_{vxww}A_{xpyy}}{\left(\frac{1}{N}\mathrm{tr}(\overline {V^2})\right)^4}=\frac{\frac{1}{N}A_{pvuu}A_{vxww}A_{cpyy}}{\left(\frac{1}{N}\mathrm{tr}(\overline {V^2})\right)^3}
=1\nonumber\\
\frac{\frac{1}{N}\langle\mathrm{tr}
\contraction[2ex]{}{\Vk}{}{\Vk}
\contraction[2ex]{\Vk\Vk}{\Vk}{}{\Vk}
\contraction[4ex]{\Vk\Vk\Vk\Vk\Vk}{\Vk}{\Vk}{\Vk}
\contraction[2ex]{\Vk\Vk\Vk\Vk}{\Vk}{\Vk}{\Vk}
\Vk \Vk \Vk \Vk\Vk \Vk\Vk\Vk\rangle}{\left(\frac{1}{N}\mathrm{tr}(\overline {V^2})\right)^4}
&=\frac{\frac{1}{N}A_{ptqq}A_{tvuu}A_{vyxw}A_{wpyx}}{\left(\frac{1}{N}\mathrm{tr}(\overline {V^2})\right)^4}=\frac{\frac{1}{N}A_{pvuu}A_{vyxw}A_{wpyx}}{\left(\frac{1}{N}\mathrm{tr}(\overline {V^2})\right)^3}
\sim\frac{{m-k\choose k}}{{m\choose k}}\nonumber\\
\frac{\frac{1}{N}\langle\mathrm{tr}
\contraction[2ex]{}{\Vk}{}{\Vk}
\contraction[2ex]{\Vk\Vk}{\Vk}{\Vk\Vk}{\Vk}
\contraction[4ex]{\Vk\Vk\Vk}{\Vk}{\Vk\Vk}{\Vk}
\contraction[6ex]{\Vk\Vk\Vk\Vk}{\Vk}{\Vk\Vk}{\Vk}
\Vk \Vk \Vk \Vk\Vk \Vk\Vk\Vk\rangle}{\left(\frac{1}{N}\mathrm{tr}(\overline {V^2})\right)^4}
&=\frac{\frac{1}{N}A_{ptqq}A_{txwu}A_{uyxv}A_{vpyw}}{\left(\frac{1}{N}\mathrm{tr}(\overline {V^2})\right)^4}=\frac{\frac{1}{N}A_{pxwu}A_{uyxv}A_{vpyw}}{\left(\frac{1}{N}\mathrm{tr}(\overline {V^2})\right)^3}\nonumber\\
&\sim\frac{{m-k\choose k}{m-2k \choose k}}{{m\choose k}^2}
\nonumber\\\frac{\frac{1}{N}\langle\mathrm{tr}
\contraction[2ex]{}{\Vk}{}{\Vk}
\contraction[2ex]{\Vk\Vk}{\Vk}{\Vk}{\Vk}
\contraction[2ex]{\Vk\Vk\Vk \Vk \Vk}{\Vk}{\Vk}{\Vk}
\contraction[4ex]{\Vk\Vk\Vk}{\Vk}{\Vk \Vk}{\Vk}
\Vk \Vk \Vk \Vk\Vk \Vk\Vk\Vk\rangle}{\left(\frac{1}{N}\mathrm{tr}(\overline {V^2})\right)^4}
&=\frac{\frac{1}{N}A_{ptqq}A_{twvu}A_{uyxv}A_{wpyx}}{\left(\frac{1}{N}\mathrm{tr}(\overline {V^2})\right)^4}=\frac{\frac{1}{N}A_{pwvu}A_{uyxv}A_{wpyx}}{\left(\frac{1}{N}\mathrm{tr}(\overline {V^2})\right)^3}\nonumber\\
&\sim\frac{{m-k\choose k}^2}{{m\choose k}^2}.
\end{align}
Here we used the results for lower-moment diagrams given in Eqs. (\ref{trV61}),
 (\ref{trV63}), (\ref{trV64}) and (\ref{trV65}).

\subsection{Diagram $A_{putq}A_{qvut}A_{vyxw}A_{wpyx}$}
The diagram in (\ref{e}),
$4\langle\mathrm{tr}
\contraction[2ex]{}{\Vk}{\Vk}{\Vk}
\contraction[4ex]{\Vk}{\Vk}{\Vk}{\Vk}
\contraction[2ex]{\Vk \Vk \Vk \Vk}{\Vk}{\Vk}{\Vk}
\contraction[4ex]{\Vk \Vk \Vk \Vk\Vk}{\Vk}{\Vk}{\Vk}
\Vk \Vk \Vk \Vk\Vk \Vk\Vk\Vk\rangle = A_{putq}A_{qvut}A_{vyxw}A_{wpyx}$,
can also be reduced to simpler diagrams. However the mechanism at work here is slightly different, as the diagram does not involve contractions between neighbouring $V's$ and coinciding subscripts for the same $A$; it also explicitly requires the limit $l\to\infty$.

\begin{figure}[t]
\centering
(a)\includegraphics[scale=.45]{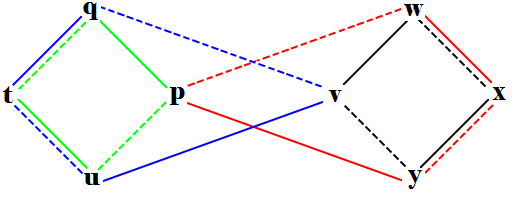}
(b)\includegraphics[scale=.45]{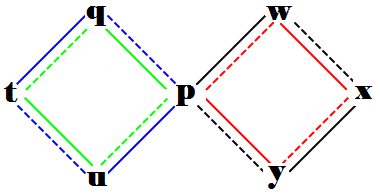}
\caption{(a) Particle diagram for $A_{putq}A_{qvut}A_{vyxw}A_{wpyx}$.
(b) Simplified diagram for  $|v\rangle=|p\rangle$, determining the contribution in  the limit $l\to\infty$. The simplified diagram  involves two standard diagrams with a shared center node $|p\rangle$.
}
\label{fig:standard_squared}
\end{figure}

The particle diagram is depicted in Fig \ref{fig:standard_squared}a. We will show that in the limit $l\to\infty$ the only relevant case is the one where the two states in the centre of the diagram, $|p\rangle$ and $|v\rangle$, coincide. The l.h.s. of  Fig \ref{fig:standard_squared}a shows two sets of  bonds corresponding to $A_{putq}$ and $A_{qvut}$. Comparing $A_{qvut}$ to $A_{putq}$, we see that bonds $\feyn{f}$ and $\feyn{h}$ are interchanged (as in the ``standard diagram'' of subsection  \ref{standard}) and moreover $|p\rangle$ is replaced by $|v\rangle$.
Now due to the results of subsection \ref{embedded_subsection}, the union of $|t\rangle$ and $|p\rangle$ must coincide with the union of $|q\rangle$ and $|u\rangle$, and the same applies to the union of $|t\rangle$ and $|v\rangle$. Hence if $|p\rangle$ and $|v\rangle$ differ at all they can do so only in states also included in $|t\rangle$ such that their unions with $|t\rangle$ can stay the same. An analogous argument holds for the r.h.s. of  Fig \ref{fig:standard_squared}a, giving as a further condition that the states in which $|p\rangle$ and $|v\rangle$ differ must also be included in $|x\rangle$. Hence $|p\rangle$ and $|v\rangle$ may differ only if $|t\rangle$ and $|x\rangle$ overlap. However requiring such an overlap would decrease the overall number of participating single-particle states and hence the argument.

We thus conclude that in the limit $l\to\infty$ the only relevant case is $|v\rangle=|p\rangle$. The diagram for this case is depicted in Fig.  \ref{fig:standard_squared}b. We see that it falls into two ``standard diagrams''  coinciding in the state $|p\rangle$, so we can use our previous reasoning for these diagrams. Choosing the state $|p\rangle$ will give a factor of $l\choose m$. Afterwards the reasoning follows identically for the left and right sides of  Fig. \ref{fig:standard_squared}b as it does for the standard diagram (in the form presented in the paragraph following (\ref{standard_unnormalized})) so that the binomial terms for each are the same as for this diagram, giving
\begin{equation}A_{putq}A_{qvut}A_{vyxw}A_{wpyx}\sim{l \choose m}\left[{{l - m} \choose k\;\;k}{m \choose k\;k}\right]^2.
\end{equation}
In the limit $l\to\infty$ this leads to
\begin{align}
\frac{\frac{1}{N}A_{putq}A_{qvut}A_{vyxw}A_{wpyx}}{\left(\frac{1}{N}\mathrm{tr}(\overline V^2)\right)^4}\sim\frac{{m-k\choose k}^2}{{m\choose k}^2}
\end{align}
as for the diagrams in (\ref{d}).

\subsection{Intermediate result}
We thus see that the diagrams (\ref{a}-\ref{e}) considered so far altogether give a contribution
(after normalization)\begin{equation}
14+28\frac{{m-k\choose k}}{{m\choose k}}+8\frac{{m-k\choose k}{m-2k \choose k}}{{m\choose k}^2}+28\frac{{m-k\choose k}^2}{{m\choose k}^2}.
\end{equation}
The remaining six diagrams from (\ref{f}) can be written in terms of $A$ as
\begin{align}
&4A_{putq}A_{qwvt}A_{upyx}A_{xywv}+
2 A_{pxwq}A_{qwvt}A_{tpyu}A_{uyxv}+ 4 A_{pxwq}A_{qwvt}A_{tyxu}A_{upyv}\nonumber\\
&8 A_{putq}A_{qwvt}A_{uyxv}A_{wpyx}
+ 8 A_{putq}A_{qxwt}A_{uyxv}A_{vpyw}
 +A_{pwvq}A_{qxwt}A_{tyxu}A_{upyv}
\end{align}
We will use the full machinery of the path summation method in the first
three cases; in the remaining cases we will exploit some features of the diagrams to simplify our calculations.

\subsection{Diagram ${A_{putq}A_{qwvt}A_{upyx}A_{xywv}}$}
\label{cube}
\begin{figure}[b]
\centering
\includegraphics[scale=0.5]{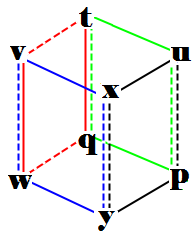}
\caption{Particle diagram for $A_{putq}A_{qwvt}A_{upyx}A_{xywv}$.}
\label{fig:cuboid}
\end{figure}

\begin{figure}[ht!]

\minipage{0.33\textwidth}
\centering
(a)\includegraphics[scale=0.25]{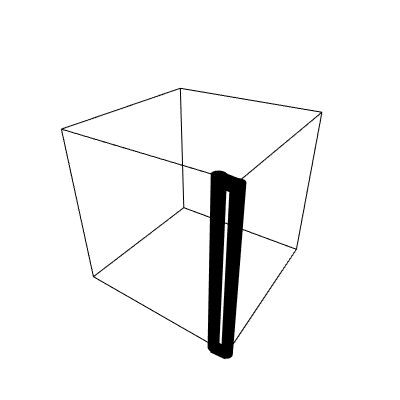}
\endminipage\hfill
\minipage{0.33\textwidth}
\centering
(b)\includegraphics[scale=.25]{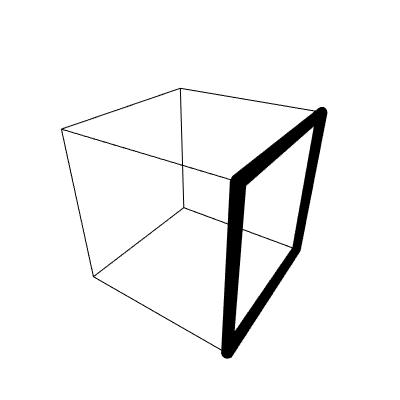}
\endminipage\hfill
\minipage{0.33\textwidth}
\centering
(c)\includegraphics[scale=0.25]{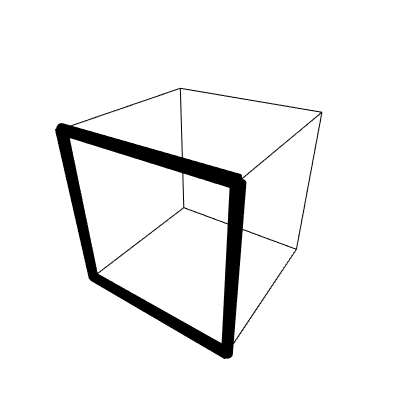}
\endminipage\hfill

\minipage{0.33\textwidth}
\centering
(d)\includegraphics[scale=.25]{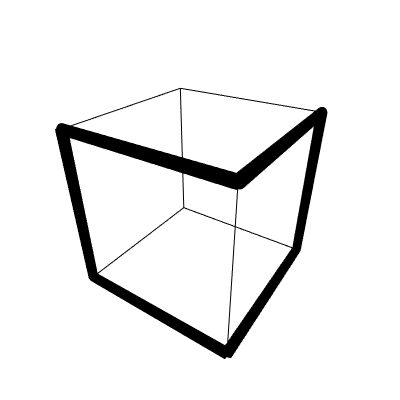}
\endminipage\hfill
\minipage{0.33\textwidth}
\centering
\centering
(e)\includegraphics[scale=0.25]{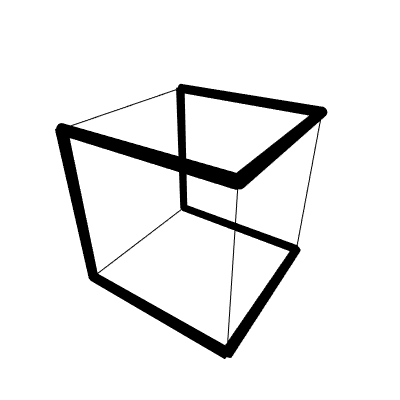}
\endminipage\hfill
\minipage{0.33\textwidth}
\centering
(f)\includegraphics[scale=.25]{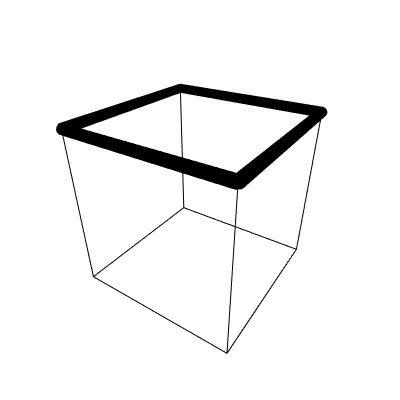}
\endminipage\hfill
\caption{ Illustrations of the loops
 (a) $\protect\overrightarrow{xy}$, (b) $\protect\overrightarrow{xupy}$,
(c) $\protect\overrightarrow{vxyw}$, (d) $\protect\overrightarrow{vxupyw}$, (e) $\protect\overrightarrow{vxutqpyw}$,
(f) $\protect\overrightarrow{tvxu}$.
By maximising the argument we find that the loops in (a) and (f) have to contain $k$ single-particle states. The remaining loops are included in table (\ref{table}). The number of single-particle states in these loops is not uniquely determined by maximising the argument.}
\label{loops}
\end{figure}
The particle diagram ${A_{putq}A_{qwvt}A_{upyx}A_{xywv}}$
has a cube structure as shown in Fig. \ref{fig:cuboid}. This diagram resembles the prism diagram from subsection \ref{prism}, in particular because the faces on the sides also correspond to the different factors $A$.
By the same arguments as for the prism there are loops (see e.g. Fig. \ref{loops}a-e)
that visit all permitted sequences of nodes on the upper face of the cube followed by a motion in the opposite direction at the bottom and closing of the loop,\begin{align}
&\arr{t\bo{q}},\arr{v\bo{w}},\arr{x\bo{y}},\arr{u\bo{p}}; \nonumber\\
&\arr{tv\bo{wq}}, \arr{vx\bo{yw}}, \arr{xu\bo{py}}, \arr{ut\bo{qp}};\nonumber\\ &\arr{tvx\bo{ywq}}, \arr{vxu\bo{pyw}}, \arr{xut\bo{qpy}}, \arr{utv\bo{wqp}};\nonumber\\ &\arr{tvxu\bo{pywq}}, \arr{vxut\bo{qpyw}}, \arr{xutv\bo{wqpy}}, \arr{utvx\bo{ywqp}}.
\end{align}
 In addition there are loops travelling only on the top face ($\overrightarrow{tvxu}$, see Fig. \ref{loops}f) or the bottom face ($\overrightarrow{qwyp}$). The numbers of states included in these loops are identical and will be denoted  by $n_{\rm top}$. This leads to the four equations
\begin{align}
\label{sys1c}
n_{tv\bo{wq}}+n_{utv\bo{wqp}}+n_{tvx\bo{ywq}}+n_{xutv\bo{wqpy}}+n_{utvx\bo{ywqp}}+n_{tvxu\bo{pywq}}+n_{\rm top}&=k\nonumber\\
n_{vx\bo{yw}}+n_{tvx\bo{ywq}}+n_{vxu\bo{pyw}}+n_{utvx\bo{ywqp
}}+n_{tvxu\bo{pywq}}+n_{vxut\bo{qpyw}}+n_{\rm top}&=m-k\nonumber\\
n_{xu\bo{py}}+n_{vxu\bo{pyw}}+n_{xut\bo{qpy}}+n_{tvxu\bo{pywq}}+n_{vxut\bo{qpyw}}+n_{xutv\bo{wqpy}}+n_{\rm top}&=m-k\nonumber\\
n_{ut\bo{qp}}+n_{xut\bo{qpy}}+n_{utv\bo{wqp}}+n_{vxut\bo{qpyw}}+n_{xutv\bo{wqpy}}+n_{utvx\bo{ywqp}}+n_{\rm top}&=m-k.
\end{align}
analogous to (\ref{sys1}). The eight bonds on the sides give rise to the equations
\begin{align}
\label{sys2c}
n_{t\bo{q}}+n_{tv\bo{wq}}+n_{tvx\bo{ywq}}+n_{tvxu\bo{pywq}}&=k\nonumber\\
n_{v\bo{w}}+n_{vx\bo{yw}}+n_{vxu\bo{pyw}}+n_{vxut\bo{qpyw}}&=m-k\nonumber\\
n_{x\bo{y}}+n_{xu\bo{py}}+n_{xut\bo{qpy}}+n_{xutv\bo{wqpy}}&=k\nonumber\\
n_{u\bo{p}}+n_{ut\bo{qp}}+n_{utv\bo{wqp}}+n_{utvx\bo{ywqp}}&=k\nonumber\\
n_{t\bo{q}}+n_{ut\bo{qp}}+n_{xut\bo{qpy}}+n_{vxut\bo{qpyw}}&=m-k\nonumber\\
n_{v\bo{w}}+n_{tv\bo{wq}}+n_{utv\bo{wqp}}+n_{xutv\bo{wqpy}}&=k\nonumber\\
n_{x\bo{y}}+n_{vx\bo{yw}}+n_{tvx\bo{ywq}}+n_{utvx\bo{ywqp}}&=k\nonumber\\
n_{u\bo{p}}+n_{xu\bo{py}}+n_{vxu\bo{pyw}}+n_{tvxu\bo{pywq}}&=k\;
\end{align}
analogous to (\ref{sys2}).
Solving this system of equations one sees
that the total number of participating states is  given by $m+2k+2n_{\rm top}$ and again maximised by choosing
\begin{align}
n_{\rm top}=k
\end{align}
As before all states in the bond  $t\feyn{h}v$ are thus
taken up by the loop on the top face and we have
\begin{equation}
n_{tv\bo{wq}}=n_{utv\bo{wqp}}=n_{tvx\bo{ywq}}=n_{xutv\bo{wqpy}}=n_{utvx\bo{ywqp}}=n_{tvxu\bo{pywq}}=0,
\end{equation}
and arguments similar to subsection \ref{prism} then also give
\begin{equation}
n_{tq}=n_{vw}=k.
\end{equation}
If we insert these results into the system of equations above it reduces to\footnote{Upon substitution all but one line of (\ref{sys1c}) and (\ref{sys2c}) become trivial or reduce to lines in (\ref{sysredc}). The only exception is the third line of (\ref{sys1c}); it reduces to $n_{xu\bo{py}}+n_{vxu\bo{pyw}}+n_{xut\bo{qpy}}+n_{vxut\bo{qpyw}}=m-2k$
which is also implied by (\ref{sysredc}).}
\begin{align}
\label{sysredc}
n_{u\bo{p}}+n_{ut\bo{qp}}&=k\nonumber\\
n_{x\bo{y}}+n_{xu\bo{py}}+n_{xut\bo{qpy}}&=k\nonumber\\
n_{vx\bo{yw}}+n_{vxu\bo{pyw}}+n_{vxut\bo{qpuw}}&=m-2k\nonumber\\
n_{x\bo{y}}+n_{vx\bo{yw}}&=k\nonumber\\
n_{u\bo{p}} +n_{xu\bo{py}}+n_{vxu\bo{pyw}}&=k\nonumber\\
n_{ut\bo{qp}}+n_{xut\bo{qpy}}+n_{vxut\bo{qpyw}}&=m-2k
\end{align}
Here the six lines respectively
fix the number of states participating in
the black bond $u\feyn{h}p$, the blue bond $x\feyn{h}y$, the red bond $v\feyn{f}w$ (excluding the $k$ states in $\overrightarrow{vw}$), the black bond $x\feyn{h}y$, the green bond $u\feyn{h}p$, and the red bond $t\feyn{f}q$ (excluding the $k$ states in $\overrightarrow{tq}$). A convenient way to visualise the equations is the following table:
\begin{equation}
\label{table}
\begin{tabular}{|c|c|c||c|}\hline
 & $n_{u\bo{p}}$ & $n_{ut\bo{qp}}$ & $k$ \\\hline
$n_{x\bo{y}}$ & $n_{xu\bo{py}}$ & $n_{xut\bo{qpy}}$ & $k$ \\\hline
$n_{vx\bo{yw}}$ & $n_{vxu\bo{pyw}}$ & $n_{vxut\bo{qpyw}}$ & $m-2k$ \\\hline\hline
$k$ & $k$ & $m-2k$ &  $m$ \\\hline
\end{tabular}
\end{equation}
Here the numbers in each row or column have to sum to the number given in the end, and the upper left corner is left empty. The form of this table will be very helpful in the following. To evaluate the diagram we have to consider all ways to select from altogether $l$ single-particle states  four loops with $k$ states each (along the sides of the top and bottom face, and the two loops $\overrightarrow{tq}$ and $\overrightarrow{vw}$) as well as $m$ states included in the loops listed in the table. This leads to a factor ${l\choose k\;k\;k\;k\;m}$. Then we divide the states contributing to loops in the table into groups contributing to the three columns; there are ${m \choose k\;k}={m\choose k}{m-k\choose k}$ ways to do this. Next we select from among the $m-k$ states associated to the second and third column $k$ states to make up the sums in the first row; this leads to a factor ${m-k\choose k}$.
We are now left with $m-k$ states making up the second and third row (those contributing to the first column and the states contributing to the other columns and not selected in the previous step). There are ${m-k\choose k}$ ways to distribute these states between the two rows in question. This leads to a result of
\begin{equation}
{A_{putq}A_{qwvt}A_{upyx}A_{xywv}}\sim{l\choose k\;k\;k\;k\;m}{m\choose k}{m-k\choose k}^3
\end{equation}
and, after normalization (using (\ref{multi2})),
\begin{equation}
\frac{\frac{1}{N}{A_{putq}A_{qwvt}A_{upyx}A_{xywv}}}{\left(\frac{1}{N}\mathrm{tr}(\overline V^2)\right)^4}\sim\frac{{m-k\choose k}^3}{{m\choose k}^3}.
\end{equation}

\subsection{Diagram $A_{pxwq}A_{qwvt}A_{tpyu}A_{uyxv}$}
\begin{figure}
\centering
\includegraphics[scale=0.5]{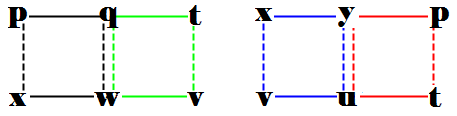}
\caption{Particle diagram for $A_{pxwq}A_{qwvt}A_{tpyu}A_{uyxv}$.}
\label{fig:standard_collapse_1}
\end{figure}

The term $A_{pxwq}A_{qwvt}A_{tpyu}A_{uyxv}$ with diagram shown in Fig. \ref{fig:standard_collapse_1} will also be calculated using path summation. Explicitly, given all permitted paths in the diagram we optimise the number of single-particle states contained in each path in order to maximise the argument. In this way we calculate all \emph{contributing} paths -- those which do not necessarily have to contain zero elements in order to maximise the argument given by the sum $\mathrm{arg} = \sum_j n_j$. The permitted closed loops for $A_{pxwq}A_{qwvt}A_{tpyu}A_{uyxv}$ are as follows
\begin{equation}\begin{array}{ll}
\alpha ~=~\overrightarrow{ptq} & \rho ~=~\overrightarrow{ptvx}\\
\beta ~=~\overrightarrow{pxy} & \xi ~=~\overrightarrow{qtpxvw} \\
\gamma ~=~\overrightarrow{xvw} & \theta ~=~\overrightarrow{qptvxw} \\
\delta ~=~\overrightarrow{vtu} & \lambda ~=~\overrightarrow{ypxvtu} \\
\epsilon ~=~\overrightarrow{qwvt} & \nu ~=~\overrightarrow{yxptvu} \\
\eta ~=~\overrightarrow{qwxp} & \pi ~=~\overrightarrow{pqtuvwxy} \\
\omega ~=~\overrightarrow{yutp} & \sigma ~=~\overrightarrow{qw} \\
\mu ~=~\overrightarrow{yuvx}  & \tau ~=~\overrightarrow{yu}.
\end{array}\end{equation}
Because the number of single-particle states contained in all loops which pass through a bond $\feyn{h}$ must sum to $k$ and similarly the number of single-particle states contained in all paths which pass through a bond $\feyn{f}$ must sum to $m-k$ we can immediately read off the following equations.
 \allowdisplaybreaks[1]
\begin{align}
\label{eq}
n_{\alpha}+n_{\xi}+n_{\theta}+n_{\nu}+n_{\rho}&=k\nonumber\\
n_{\beta}+n_{\xi}+n_{\lambda}+n_{\nu}+n_{\rho}&=k\nonumber\\
n_{\gamma}+n_{\xi}+n_{\theta}+n_{\lambda}+n_{\rho}&=k\nonumber\\
n_{\delta}+n_{\theta}+n_{\lambda}+n_{\nu}+n_{\rho}&=k\\
\nonumber\\
n_{\epsilon}+ n_{\xi}+ n_{\sigma}&=k\nonumber\\
n_{\eta}+n_{\theta}+n_{\sigma}&=k\nonumber\\
n_{\omega}+n_{\lambda}+n_{\tau}&=k\nonumber\\
n_{\mu}+n_{\nu}+n_{\tau}&=k\\
\nonumber\\
n_{\epsilon}+n_{\xi}+n_{\omega}+n_{\pi}+n_{\alpha}&=m-k\nonumber\\
n_{\epsilon}+n_{\xi}+n_{\mu}+n_{\pi}+n_{\gamma}&=m-k\nonumber\\
n_{\epsilon}+n_{\omega}+n_{\lambda}+n_{\pi}+n_{\delta}&=m-k\nonumber\\
n_{\epsilon}+n_{\mu}+n_{\nu}+n_{\pi}+n_{\delta}&=m-k\\
\nonumber\\
n_{\eta}+n_{\theta}+n_{\omega}+n_{\pi}+n_{\alpha}&=m-k\nonumber\\
n_{\eta}+n_{\omega}+n_{\lambda}+n_{\pi}+n_{\beta}&=m-k\nonumber\\
n_{\eta}+n_{\mu}+n_{\nu}+n_{\pi}+n_{\beta}&=m-k\nonumber\\
n_{\eta}+n_{\theta}+n_{\mu}+n_{\pi}+n_{\gamma}&=m-k\;.
\end{align}
Using these equations the argument, i.e. the sum of all $n_j$'s can be written as
\begin{equation}
\arg=m+2k+2n_\delta-2n_\xi.
\end{equation} To maximise this number we have to choose
\begin{equation}
n_\delta=k,\quad n_\xi=0
\end{equation}
which leads to the value $\arg=m+4k$ expected due to our considerations in
subsection \ref{arguments}. Substituting this result back into the system of equations yields
\begin{align}
&n_{\alpha} ~=~n_{\beta} ~=~n_{\gamma} ~ ~=~k \nonumber\\
&n_{\theta} ~=~n_{\lambda} ~=~n_{\nu} ~=~n_{\rho} ~=~0 \nonumber\\
&n_{\eta} ~=~n_{\epsilon} \nonumber\\
&n_{\mu} ~=~n_{\omega} \nonumber\\
&n_{\sigma} = k-n_{\epsilon} \nonumber\\
&n_{\tau} = k-n_{\omega}\nonumber\\
&n_{\pi} =m-2k-n_{\epsilon}-n_{\omega}
\end{align}
where the additional zeros follow from the fourth line of (\ref{eq}) and
$n_\delta=k$. As the solutions are parametrised by the variables $n_\epsilon$
and $n_\omega$ we obtain a sum over
multinomial contributions as in
(\ref{multi}), \begin{align}\label{eq:hahn_term}
&A_{pxwq}A_{qwvt}A_{tpyu}A_{uyxv} \sim \sum_{n_{\epsilon},n_{\omega}}{{l}\choose{n_{\alpha}~n_{\beta}~n_{\gamma}~n_{\delta}~n_{\epsilon}~n_{\sigma}~n_{\omega}~n_{\tau}~n_{\eta}~n_{\omega}~n_{\pi}}}\nonumber\\
&= \sum_{n_{\epsilon},n_{\omega}}{l \choose {k~k~k~k~n_{\epsilon}~k-n_\epsilon~n_\omega~k-n_\omega~n_\epsilon~n_\omega~m-2k-n_\epsilon-n_\omega
~}}\nonumber\\
&=\sum_{n_\epsilon,n_\omega}{l \choose k~k~k~k~k~k~m-2k}{k\choose n_\epsilon}{k\choose
n_\omega}{m-2k\choose n_\epsilon}{m-2k-n_\epsilon\choose n_\omega}\nonumber\\
&={l \choose k~k~k~k~k~k~m-2k}\sum_{n_\epsilon}{k\choose n_\epsilon}{m-2k\choose n_\epsilon}{m-k-n_\epsilon\choose k}\;.
\end{align}
Here in the final step we used the combinatorial identity
\begin{equation}
\label{identity}
\sum_c{a\choose c}{b\choose c}={a+b \choose a}
\end{equation}
(a special case of Vandermode's identity) to perform the sum over $n_\omega$.
Including the normalization and using (\ref{multi2}) we then attain
\begin{equation}
\frac {{\frac{1}{N}}{A_{pxwq}A_{qwvt}A_{tpyu}A_{uyxv}}} { \left({\frac{1}{N}} \mathrm{tr}(\overline{V_{k}^2})\right)^4 }
 = \frac{{{m-k}\choose{k}}}{{{m\choose {k}}}^3} {\sum_{n_{\epsilon}} {{m-k-n_{\epsilon}}\choose{k}} {{m-2k}\choose{n_{\epsilon}}} {{k}\choose{n_{\epsilon}}}}.
\label{eq:hahn_poly}
\end{equation}
Extraordinarily the final result still contains a sum over $n_\epsilon$,
and this sum is proportional to a \textit{Hahn polynomial}.
This is an interesting development because it hints that we can perhaps make further progress with a method based on the structure of such polynomials. Due to the factor ${m-2k\choose n_\epsilon}$ the present diagram contributes only for $m\geq 2k$ (initially only with the summand $n_\epsilon=0$).
 \\
\subsection{Hahn Polynomials}
We briefly highlight the following lemma to express (\ref{eq:hahn_poly}) in the form as shown in \cite{small}. The proof uses the formalism of the class of \textit{Hahn polynomials} for which definitions and further details can be found in part 9.5 of the compendium \cite{koekoek}. For a more detailed proof see \cite{thesis}.\\ \ \\
\noindent\textit{Lemma.}
\begin{equation}{\frac{{{m-k}\choose{k}}}{{{{m}\choose{k}}^3}}}{\sum_{n_{\epsilon}} {{m-k-n_{\epsilon}}\choose{k}}{{k}\choose{n_{\epsilon}}}{{m-2k}\choose{n_{\epsilon}}}} = {\frac{{{{m-k}\choose{k}}^2}}{{{{m}\choose{k}}^3}}} \sum_p \frac{{{{{k}\choose {p}}^2}{{m-2k}\choose {k-p}}}}{{{m-k}\choose{p}}}\\ \end{equation}

\noindent\textit{Proof (Outline)}.\\

Writing the l.h.s as an hypergeometric function
\begin{align}&{\sum_{n_{\epsilon}} {{m-k-n_{\epsilon}}\choose{k}}{{k}\choose{n_{\epsilon}}}{{m-2k}\choose{n_{\epsilon}}}}\nonumber\\ &= {{m-k}\choose{k}} {~}_3 F_2\left( \begin{array}{cc}
-k, 2k-m, 2k-m\\
1, k-m \end{array} ;1\right)  \end{align}
and recalling the definition of a Hahn polynomial
\begin{equation}Q_n (x; \alpha, \beta, N) := {~}_3 F_2\left( \begin{array}{cc}
-n, n+\alpha+\beta+1,-x\\
\alpha+1, -N \end{array} ;1\right) \end{equation}
we have $n=k$, $\alpha=0$, $\beta = k-m-1$ and $x=m-2k$. To express this as a series it should then be noted that
\begin{equation}
{~}_1 F_1\left( \begin{array}{cc}
-x\\
\alpha+1\end{array} ;-t\right)  {~}_1 F_1\left( \begin{array}{cc}
x-N\\
\beta+1\end{array} ;t\right) = \sum_{n=0}^N{\frac{(-N)_n}{(\beta+1)_nn!}} Q_{n}(x; \alpha, \beta, N) t^n.
 \end{equation}
Further, using the series expansion of ${}_1F_1$ it can be shown that this
product is \begin{equation}{~}_1 F_1\left( \begin{array}{cc}
2k-m\\
1\end{array} ;-t\right) {~}_1 F_1\left( \begin{array}{cc}
{-k}\\
{-(m-k)}\end{array} ;t\right) = \sum_{n,p=0}^{\infty} \frac{{{m-2k}\choose n}{k\choose p}}{{{m-k}\choose p}}\frac{{t^{n+p}}}{{n!p!}}.\end{equation}
Comparing the coefficients of $t^k$ thus gives
 \begin{equation}Q_k(x; \alpha, \beta, N) = \sum_{p=0}^{k} \frac{{{{{k}\choose {p}}^2}{{m-2k}\choose {k-p}}}}{{{m-k}\choose{p}}} \end{equation}
which completes the proof.\hfill$\square$
\subsection{Diagram $A_{pxwq}A_{qwvt}A_{tyxu}A_{upyv}$}
\begin{figure}[t]
\centering
\includegraphics[scale=0.5]{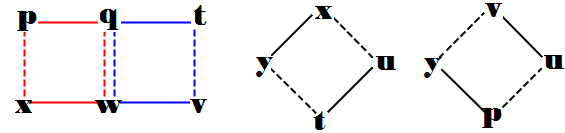}
\caption{Particle diagram for  $A_{pxwq}A_{qwvt}A_{tyxu}A_{upyv}$.}
\label{fig:diamond_collapse_1}
\end{figure}

 Identifying loops and maximising the argument for $A_{pxwq}A_{qwvt}A_{tyxu}A_{upyv}$ with corresponding diagram Fig. \ref{fig:diamond_collapse_1} in the same manner as described previously we attain the following loops
 [1]
\begin{equation}\begin{array}{ll}
\alpha ~=~\overrightarrow{pqwxy} & \sigma ~=~\overrightarrow{uxwqputv}\\
\beta ~=~\overrightarrow{qtuvw} & \tau ~=~\overrightarrow{uvwqtypx}\\
\gamma ~=~\overrightarrow{pyx} & \phi ~=~\overrightarrow{utqwvyxp}\\
\delta ~=~\overrightarrow{tuv} & \theta ~=~\overrightarrow{ytqwv} \\
\epsilon ~=~\overrightarrow{pqty} & \lambda ~=~\overrightarrow{upqwx} \\
\eta ~=~\overrightarrow{xwvy} & \nu ~=~\overrightarrow{ytv} \\
\omega ~=~\overrightarrow{vwxu} & \pi ~=~\overrightarrow{upx} \\
\mu ~=~\overrightarrow{tqpu} & \psi ~=~\overrightarrow{ypxyvt}\\
\xi ~=~\overrightarrow{ypqtuvwx} & \kappa ~=~\overrightarrow{utvyxp}\\
\rho ~=~\overrightarrow{ypqwxuvt} & \chi ~=~\overrightarrow{qw}
\end{array}\end{equation}
and the sizes of the bonds are constrained by the following equations [1]
\begin{align}
n_{\alpha}+n_{\lambda}+n_{\sigma}+n_{\rho}+n_{\chi}&=k\nonumber\\
n_{\beta}+n_{\theta}+n_{\tau}+n_{\phi}+n_{\chi}&=k\\
\nonumber\\
\label{ks}
n_{\epsilon}+n_{\theta}+n_{\nu}+n_{\rho}+n_{\tau}+n_{\psi}&=k\nonumber\\
n_{\eta}+n_{\theta}+n_{\nu}+n_{\sigma}+n_{\phi}+n_{\kappa}&=k\nonumber\\
n_{\omega}+n_{\lambda}+n_{\pi}+n_{\rho}+n_{\tau}+n_{\psi}&=k\nonumber\\
n_{\mu}+n_{\lambda}+n_{\pi}+n_{\sigma}+n_{\phi}+n_{\kappa}&=k\nonumber\\
n_{\gamma}+n_{\pi}+n_{\tau}+n_{\phi}+n_{\psi}+n_{\kappa}&=k\nonumber\\
n_{\delta}+n_{\nu}+n_{\rho}+n_{\sigma}+n_{\psi}+n_{\kappa}&=k\\
\nonumber\\
n_{\alpha}+n_{\gamma}+n_{\epsilon}+n_{\xi}+n_{\rho}+n_{\tau}+n_{\psi}&=m-k\nonumber\\
n_{\alpha}+n_{\gamma}+n_{\eta}+n_{\xi}+n_{\sigma}+n_{\phi}+n_{\kappa}&=m-k\nonumber\\
n_{\alpha}+n_{\epsilon}+n_{\mu}+n_{\xi}+n_{\lambda}+n_{\rho}+n_{\sigma}&=m-k\nonumber\\
n_{\alpha}+n_{\eta}+n_{\omega}+n_{\xi}+n_{\lambda}+n_{\rho}+n_{\sigma}&=m-k\\
\nonumber\\
n_{\beta}+n_{\epsilon}+n_{\mu}+n_{\xi}+n_{\theta}+n_{\tau}+n_{\phi}&=m-k\nonumber\\
n_{\beta}+n_{\delta}+n_{\mu}+n_{\xi}+n_{\sigma}+n_{\phi}+n_{\kappa}&=m-k\nonumber\\
n_{\beta}+n_{\delta}+n_{\omega}+n_{\xi}+n_{\rho}+n_{\tau}+n_{\psi}&=m-k\nonumber\\
n_{\beta}+n_{\eta}+n_{\omega}+n_{\xi}+n_{\theta}+n_{\tau}+n_{\phi}&=m-k
\end{align}
Using these equations the argument can be written as
\begin{equation}\mathrm{arg} = \sum_j n_j\nonumber= m+2k+n_\mu+n_\eta-2n_\psi-2n_\tau-2n_\rho
\end{equation}
which reaches its maximal value $m+4k$   if
\begin{equation}
n_\mu=n_\eta=k,\quad\quad n_\psi=n_\tau=n_\rho=0.
\end{equation}
This gives the following restrictions on the number of single-particle states in  the loops
\begin{align}
n_{\theta} &~=~n_{\lambda }~=~n_{\nu }~=~n_{\pi }~=~n_{\sigma }~=~n_{\phi }~=~n_{\kappa }~=~0\nonumber\\
n_{\gamma} &~=~n_{\delta }~=~n_{\epsilon }~=~n_{\omega }~=~k\nonumber\\
n_{\beta }&~=~n_{\alpha }\end{align}
(where
the additional zeros are due to the second and fourth member of (\ref{ks}) and $n_\mu=n_\eta=k$)
as well as the  identities
\begin{align}
n_{\chi} &= k-n_{\alpha}\nonumber\\
n_{\xi} &= m-3k-n_{\alpha}.
\end{align}
We have to sum over all possible values for the remaining parameter $n_\alpha$,
and for each of these the number of ways to select the single-particle states
participating in the loops is given by a multinomial. This leads to
\begin{align}&A_{pxwq}A_{qwvt}A_{tyxu}A_{upyv} \sim \sum_{n_{\alpha}}{l \choose {n_{\mu}~n_\eta~n_{\gamma}~n_\delta~n_\epsilon~n_\omega~n_{\alpha}~n_{\chi}~n_{\beta}~n_{\xi}}}\nonumber\\
&=\sum_{n_{\alpha}}{l \choose {k~k~k~k~k~k~n_{\alpha}~~k-n_{\alpha}~n_{\alpha}~m-3k-n_{\alpha}}}\nonumber\\
&=\sum_{n_{\alpha}}{l \choose {k~k~k~k~k~k~~k~~m-3k}}{k \choose n_\alpha}{m-3k
\choose n_\alpha}\nonumber\\
&={l \choose {k~k~k~k~k~k~~k~~m-3k}}{m-2k\choose k}
\end{align}
where in the last step the sum over $n_\alpha$ was evaluated using (\ref{identity}).
Using (\ref{multi2}) the normalised result is then obtained as
\begin{align}\frac{{\frac{1}{N}}A_{pxwq}A_{qwvt}A_{tyxu}A_{upyv}}{\left({\frac{1}{N}} \mathrm{tr}(\overline{V_{k}^2})\right)^4 } \sim \frac{{{m-k}\choose{k}}{{m-2k}\choose{k}}^2}{{m\choose k}^3}.
\end{align}

\subsection{Diagram $A_{putq}A_{qwvt}A_{uyxv}A_{wpyx}$}
\label{renorm}
\begin{figure}[t]
\centering
\includegraphics[scale=0.5]{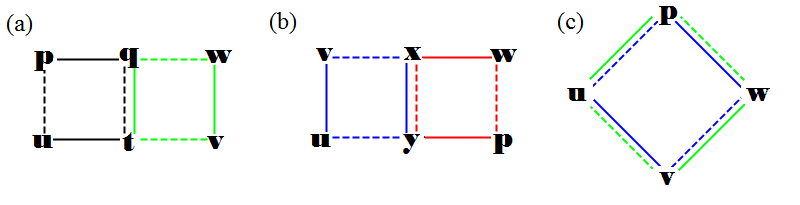}
\caption{Particle diagram of the term $A_{putq}A_{qwvt}A_{uyxv}A_{wpyx}$.   (a) and (b) depict the parts associated to $A_{putq}A_{qwvt}$ and $A_{uyxv}A_{wpyx}$ respectively. (c) depicts the relationship between the many-particle states appearing as corners in (a) and (b), with the outer green square arising from (a) and the inner blue square arising from (b). (c) coincides with the standard diagram.}
\label{fig:squared}
 \end{figure}

For the remaining diagrams we will use some interesting shortcuts. The evaluation of the diagram $A_{putq}A_{qwvt}A_{uyxv}A_{wpyx}$ can be reduced to the ``standard diagram'' using a process resembling a renormalization procedure. Let us first consider the particle diagram just representing the first two factors $A_{putq}A_{qwvt}$, shown in Fig. \ref{fig:squared}a.
In this diagram the states in the corners  $|p\rangle$, $|w\rangle$, $|u\rangle$, $|v\rangle$  have only two bonds attached; these bonds belong to the same factor $A$ and hence cannot form part of the same path. Hence the appropriate paths to consider in this reduced diagram are not only the closed loop $\overrightarrow{qt}$ but also paths that are not closed but instead begin and end at the corner nodes.
By analogy  e.g. to subsection \ref{standard}, it is natural to assume that the bonds $q\feyn{f}t$ and $q\feyn{h}t$ should overlap maximally, implying that for the case $m>2k$ all $k$ states of $q\feyn{h}t$ are also included in $q\feyn{f}t$ and thus in the loop $\overrightarrow{qt}$.
We omit a detailed proof of this assumption but it can be verified \cite{thesis} by keeping the number of states in $\overrightarrow{qt}$ as a variable in the following calculations and then choosing the value of this variable in the end to maximise the argument. The remaining $m-2k$ states from the bond $q\feyn{f}t$ can only form part of the open path $p\feyn{f}q\feyn{f}t\feyn{f}u$, as all other options would involve subsequent bonds associated to the same factor $A$. Together with the bond $p\feyn{h}u$ this leads to an overlap of $m-k$ states between $|p\rangle$ and $|u\rangle$. The remaining $k$ states in $p\feyn{f}q$ must coincide with $q\feyn{h}w$ forming the open path $p\feyn{f}q\feyn{h}w$. An analogous argument leads to $k$ states in the path $u\feyn{f}t\feyn{h}v$. Together with the bond $w\feyn{f}v$ of $m-k$ states the overlaps between the corner states are thus just as depicted in the outer blue square in Fig. \ref{fig:squared}c, coinciding with the conditions that usually arise from a single factor $A$. Again neighbouring bonds in this effective diagram may not share any single particle states.\footnote{
Consider e.g. the paths attached to $|p\rangle$ in Fig. \ref{fig:squared}a, $p\feyn{h}u$, $p\feyn{f}q\feyn{f}t\feyn{f}u$, and $p\feyn{f}q\feyn{h}w$. These paths must be disjoint due to the arguments in subsection \ref{embedded_subsection}. Then the same must apply to the two outer (green) bonds attached to $|p\rangle$ in Fig. \ref{fig:squared}c, since the only difference between these bonds and the aforementioned paths is that  $p\feyn{h}u$ and $p\feyn{f}q\feyn{f}t\feyn{f}u$ were merged.} Analogous arguments for the diagram associated to $A_{uyxv}A_{wpyx}$ as depicted in Fig. \ref{fig:squared}b yield
connections according to the inner (blue) square in Fig. \ref{fig:squared}c.

Now Fig. \ref{fig:squared}c is just the standard diagram of subsection \ref{standard}, and we have already evaluated its contribution (for $m>2k$) as
${l \choose k\;k\;k\;k\;k\;m-2k}$.
However in addition to subsection \ref{standard} we also have to take into account that there are ${m-k\choose k}$ ways to distribute the $m-k$ states of the bond $p\feyn{f}u$ in Fig \ref{fig:squared}c among the paths $p\feyn{h}u$ and $p\feyn{f}q\feyn{f}t\feyn{f}u$
in Fig. \ref{fig:squared}a. Furthermore we have to choose the $k$ states included in the loop $\overrightarrow{ec}$. These can be taken from any of the $l$ single-particle states apart from the $m$ states included in other paths that visit $|q\rangle$ and/or $|t\rangle$;  hence there are ${l-m\choose k}$ possible choices. Two analogous factors arise from Fig. \ref{fig:squared}b. We obtain
\begin{equation}\label{standard_sq}
A_{putq}A_{qwvt}A_{uyxv}A_{wpyx}\sim{l \choose k\;k\;k\;k\;m-2k}{m-k\choose k}^2{l-m\choose k}^2
\end{equation}
and thus including normalization
\begin{equation}
\frac{\frac{1}{N}A_{putq}A_{qwvt}A_{uyxv}A_{wpyx}}{{\left(\frac{1}{N}\mathrm{tr}(\overline
V^2)\right)^4}}
\sim
\frac{{m-k \choose k}^3}{{m\choose k}^3}.
\label{standard_sq_res}
\end{equation}
Analogous reasoning gives a  vanishing result for $m<2k$, again in line with
(\ref{standard_sq_res}).
\subsection{Diagram $A_{putq}A_{qxwt}A_{uyxv}A_{vpyw}$} \begin{figure}[t]
\centering
\includegraphics[scale=0.5]{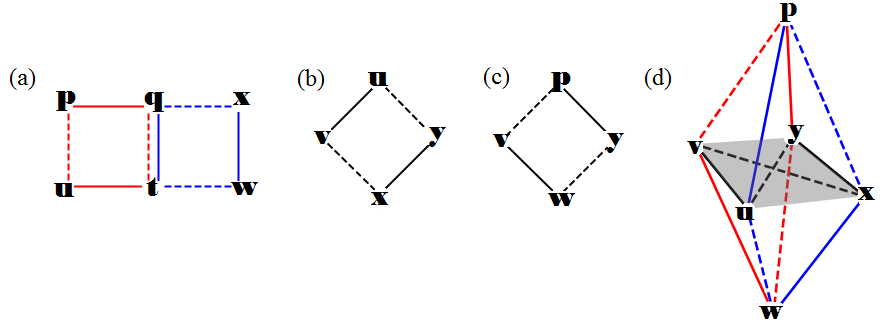}
\caption{Particle diagram for $A_{putq}A_{qxwt}A_{uyxv}A_{vpyw}$, representing the factors (a) $A_{putq}A_{qxwt}$, (b) $A_{uyxv}$, and (c) $A_{vpyw}$. Using the fact that the corner states in (a) form a square as in Fig. \ref{fig:square} and combining this square with (b) and (c) one obtains the diamond-shaped diagram in (d). The latter diagram is analogous to Fig. \ref{fig:diamond}.}
\label{fig:diamond_collapse_2}
\end{figure}
We argue similarly for the particle diagram  $A_{putq}A_{qxwt}A_{uyxv}A_{vpyw}$ as shown in Fig. \ref{fig:diamond_collapse_2}. The bonds representing the first two factors are displayed in  Fig.
\ref{fig:diamond_collapse_2} and have the same form as in Fig. \ref{fig:squared}. Following the reasoning
of subsection \ref{renorm} one sees that the corner points $|p\rangle$, $|x\rangle$, $|u\rangle$, and $|w\rangle$ can be regarded as forming a square, as arising from a single factor $A$. Together with the squares from the third and fourth factor displayed in Figs. \ref{fig:diamond_collapse_2}b and  \ref{fig:diamond_collapse_2}c we obtain  a diamond-shaped diagram as in Fig. \ref{fig:diamond_collapse_2}d. This diagram coincides with the one from subsection \ref{diamond} and displayed in Fig. \ref{fig:diamond} up to renaming of nodes.
If we count the participating single-particle states ignoring the nodes $|q\rangle$ and $|t\rangle$ in the center of Fig. \ref{fig:diamond_collapse_2}a we obtain the same result as earlier (see (\ref{eq:sixth_sr})),  ${l\choose k\;k\;k\;k\;k\;k\;m-3k}$.
In analogy to subsection \ref{renorm} this has to be multiplied with a factor ${m-k\choose k}$ counting the number of ways in which the states in the ``effective'' bond $p\feyn{f}u$ in Fig. \ref{fig:diamond_collapse_2}d can be distributed among the bond $p\feyn{h}u$ in  Fig. \ref{fig:diamond_collapse_2}a and the open path $p\feyn{f}q\feyn{f}t\feyn{f}u$. In addition a factor ${l-m\choose k}$ counts the possible choices for the $k$ particles forming the loop $\overrightarrow{qt}$. Altogether this leads to
\begin{equation}
A_{putq}A_{qxwt}A_{uyxv}A_{vpyw}\sim
{l\choose
k\;k\;k\;k\;k\;k\;m-3k}{m-k\choose k}{l-m\choose k}.
\end{equation}
After normalization we obtain
\begin{equation}\frac{{\frac{1}{N}}A_{putq}A_{qxwt}A_{uyxv}A_{vpyw}}{\left({\frac{1}{N}} \mathrm{tr}(\overline{V_{k}^2})\right)^4 } \sim \frac{{{m-k}\choose{k}}^2{{m-2k}\choose{k}}}{{m\choose k}^3}.
\end{equation}

\subsection{Diagram $A_{pwvq}A_{qxwt}A_{tyxu}A_{upyv}$}
\begin{figure}[t]
\centering
\includegraphics[scale=0.5]{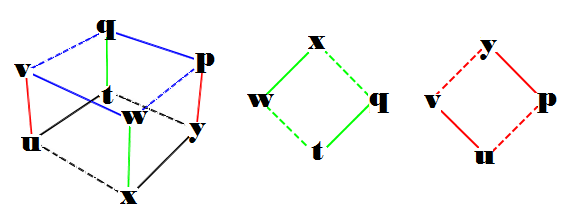}
\caption{Particle diagram for $A_{pwvq}A_{qxwt}A_{tyxu}A_{upyv}$.}
\label{fig:box}
\end{figure}
For the evaluation of the final term $A_{pwvq}A_{qxwt}A_{tyxu}A_{upyv}$ one can use the same method as outlined in the evaluation of $A_{pvuq}A_{qwvt}A_{tpwu}$ (Fig. \ref{fig:diamond}) in subsection \ref{diamond}. We first  note that the states $|v\rangle$ and $|p\rangle$ together determine all other states in the diagram, Fig \ref{fig:box}. These states are diametrically opposed in two of the squares shown in Fig. \ref{fig:box}; by the arguments in subsection \ref{embedded_subsection} they thus determine the remaining states in these squares, $|q\rangle$, $|w\rangle$, $|y\rangle$, and $|u\rangle$.
Applying the same argument again for the two remaining squares we see that the states $|x\rangle$ and $|t\rangle$ also have to be included in $|v\rangle$ and/or $|p\rangle$. Hence to maximise the argument of the diagram we minimise the overlap between $|v\rangle$ and $|p\rangle$, i.e., we include as many states as possible in loops that contain one of the nodes $|v\rangle$ or $|p\rangle$ but not both. The relevant loops are
\begin{align}\overrightarrow{vwxu},\;
\overrightarrow{vwxy},\overrightarrow{vqtu},\;
\overrightarrow{vwxqtu}
\end{align}
for $|v\rangle$ and
\begin{align}
\overrightarrow{pyxw},\;
\overrightarrow{pytq},\;
\overrightarrow{putq},\;
\overrightarrow{pyxwtq}
\end{align}
for $|p\rangle$.
If we have $m\geq 4k$ then each of these loops can be occupied by the maximal number of $k$ states. This fixes all bonds $\feyn{h}$ and $3k$ of the states in each bond $\feyn{f}$.  The only loop that can contain the remaining single-particle states is the loop $\overrightarrow{tqpyxwvu}$ formed out of all bonds $\feyn{f}$; this loop must therefore
have $m-4k$ associated single particle states. As a consequence we have
 \begin{equation}A_{pwvq}A_{qxwt}A_{tyxu}A_{upyv} \sim {l\choose{k~k~k~k~k~k~k~k~m-4k}}
\end{equation}
and thus (using (\ref{multi2}))
\begin{equation}
\label{final_diag}\frac{{\frac{1}{N}}A_{pwvq}A_{qxwt}A_{tyxu}A_{upyv}}{\left({\frac{1}{N}}
\mathrm{tr}(\overline{V_{k}^2})\right)^4 } \sim \frac{{{m-k}\choose{k}}{{m-2k}\choose{k}}{{m-3k}\choose{k}}}{{m\choose
k}^3}.
\end{equation}
In the case $m<4k$ the maximal possible number of states $2m$ remains below the value $m+4k$ required for a contribution that survives the limit $l\to\infty$;
in analogy to earlier cases this situation is accounted for correctly in (\ref{final_diag}).
\begin{figure}[H]
\centering
\includegraphics[scale=.5]{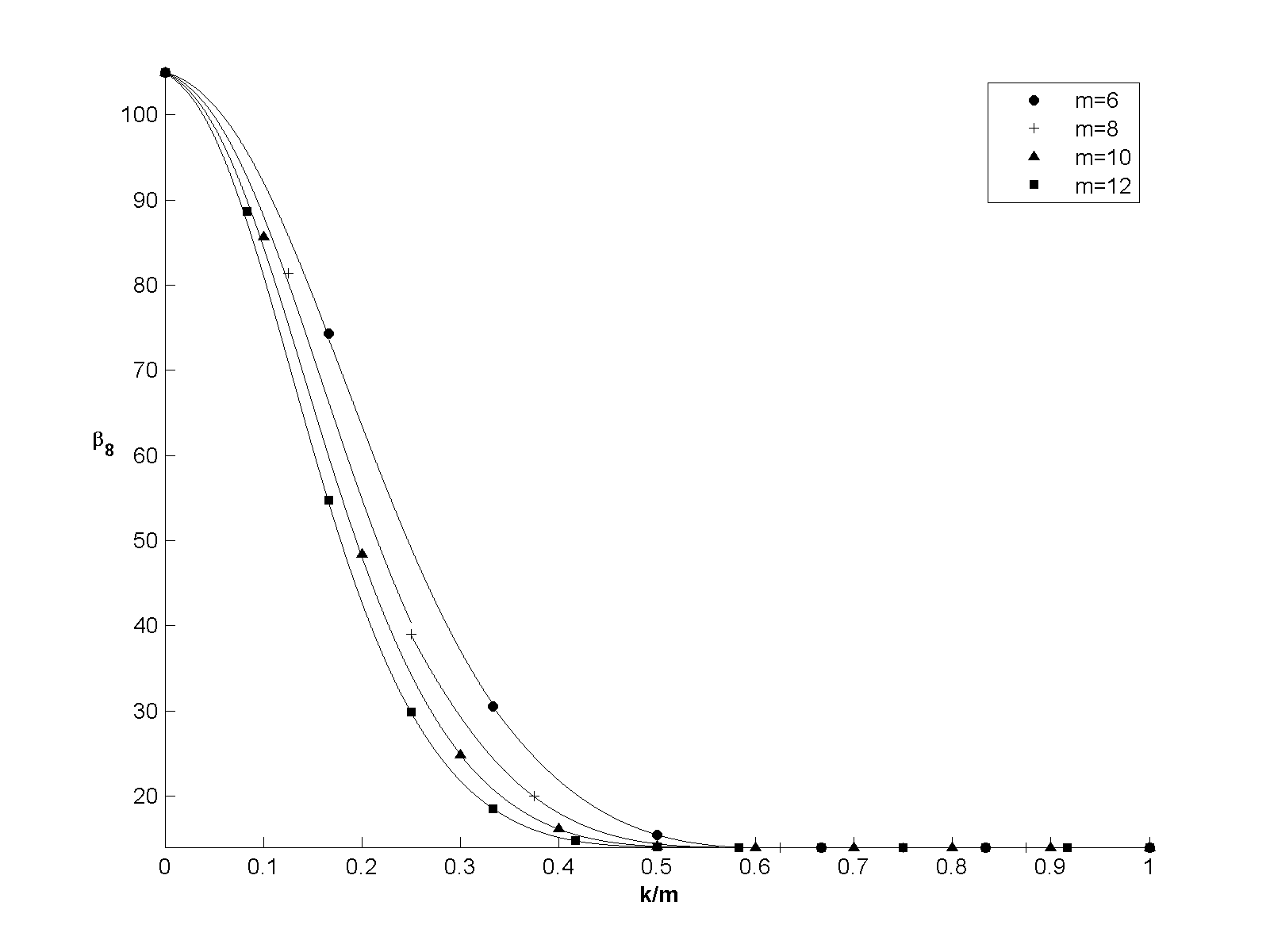}
\caption{The eighth moment $\tau$ of the level density against $k/m$ for $m=6, 8, 10, 12$ exhibiting the same properties as the fourth and sixth; a semi-circular domain, $\tau =14$ for $k/m > \frac{1}{2}$ converging to a Gaussian value, $\tau=105$ for $k \ll m \to \infty$. Higher values of $m$ give faster convergence to the semi-circular moment.}
\label{fig:eighth_moment}
\end{figure}

\subsection{Final result}
Putting it all together gives
\begin{align}\label{eq:s11}
\tau\sim&
 14 +28\frac{{{m-k}\choose{k}}}{{m\choose k}}+ 8\frac{{{m-k}\choose{k}}{{m-2k}\choose{k}}}{{m\choose k}^2} +28\frac{{{m-k}\choose{k}}^2}{{m\choose k}^2}\nonumber \\
&+12\frac{{{m-k}\choose{k}}^3}{{m\choose k}^3}+  2 \frac{{{m-k }\choose{k}}}{{{m\choose {k}}}^3} {\sum_{\alpha} {{m-k-\alpha}\choose{k}} {{m-2k}\choose{\alpha}} {{k}\choose{\alpha}}}+ 4\frac{{{m-k}\choose{k}}{{m-2k}\choose{k}}^2}{{m\choose k}^3}\nonumber\\
&+ 8\frac{{{m-k}\choose{k}}^2{{m-2k}\choose{k}}}{{m\choose k}^3} + \frac{{{m-k}\choose{k}}{{m-2k}\choose{k}}{{m-3k}\choose{k}}}{{m\choose k}^3}  \end{align}
where the term with coefficient 12 combines the results from subsections
\ref{cube} and \ref{renorm}.
Again we observe a transition between semi-circular and Gaussian behaviour,
see Fig. \ref{fig:eighth_moment}.

\section{Conclusions}
We have introduced  Feynman-like \textit{particle diagrams} as well as asymptotic approximations in the limit $l\to\infty$ to calculate statistical properties of embedded $k$-body random matrix potentials. We have illustrated the general method by confirming the known expression for the fourth moment of the average level density for the embedded GUE, and demonstrated its strength by calculating the previously inaccessible sixth and eighth moments. We have shown that in the limit $l\to\infty$ the moments for fermions and bosons have to coincide. We have also illustrated the process of \textit{path summation} and shown how this can be used to systematically implement these new techniques, even for complex
diagrams.

The results reveal that certain behaviours identified with the fourth moment follow to at least the eighth moment and plausibly to all higher moments, in particular a transition from a semi-circular moment ($\kappa = 2, h=5, \tau = 14$) for $m<2k$ to a Gaussian moment ($\kappa = 3, h=15, \tau = 105$) in
the limit $k\ll m\ll l$. We have shown that the domain of the $2n$-th moment manifests an interesting feature, namely a natural division at the points $m=2k , 3k, \ldots, nk$. For $m=k$, which is contained in the semi-circular regime,  we regain canonical RMT results as is required. In this case the bonds $\feyn{f}$ containing $m-k$ single-particle states become irrelevant, and in this context the particle diagrams can be related to Dyck Paths, Catalan Numbers and other diagrammatic methods such as \cite{kreweras} used for these systems\cite{thesis}.

Having greatly simplified  the implementation of the limit $l\to\infty$ the present ideas should be helpful for investigating further properties of  embedded  random-matrix theory in that limit, such as
the full average level density including even higher moments, spectral fluctuations, and the behaviour of the orthogonal and symplectic ensembles. Since the most complicated term in the present
work  involved a Hahn polynomial  the machinery  developed
to study such polynomials \cite{koekoek} may prove fruitful for further progress. A  relevant limit that could  be implemented in a similar way is the dense limit of bosonic systems; in this limit the
individual single-particle states are occupied by many particles such that the number of particles $m$ greatly exceeds the number of available states $l$. Another interesting question (see e.g. \cite{ullmo}) is to what extent individual complex many-body systems are faithful to random matrix averages.


\end{document}